\documentclass[aps,pre,color,psfig,epsf,notitlepage]{revtex4-1}
\usepackage{graphicx} 
\usepackage{color}
\usepackage{amsmath}
\usepackage{amsfonts}
\usepackage{amssymb}
\usepackage{enumitem}
\usepackage{comment}

\usepackage{babel}
\usepackage{txfonts}

\usepackage{tabularx}
\usepackage[dvipsnames]{xcolor}

\usepackage{array}

\usepackage{hyperref}

\usepackage[position=top,caption=false]{subfig}

\newtheorem{definition}{Definition}
\newtheorem{theorem}{Theorem}

\begin{document}

\title{Counting unlabeled multigraphs with three nodes}

\author{Andrea Bonato$^1$}
\affiliation{$^1$ SUPA, School of Physics and Astronomy, The University of Edinburgh, Edinburgh, EH9 3FD, Scotland, United Kingdom}

\begin{abstract}

Unlabeled multigraphs have diverse applications across scientific fields, from transportation and social networks to polymer physics. In particular, multigraphs are essential for studying the relationship between the spatial organization and biological function of chromatin, which is often folded into complex polymer networks whose structure is closely tied to patterns of gene expression.
A fundamental yet challenging aspect in applying graph theory to these areas is the enumeration of multigraphs, especially under structural constraints
For example, when coupled with the statistical mechanics of polymer networks, the ability to identify traversable and connected multigraphs provides powerful tools for predicting statistically favored motifs that may arise within chromatin networks.
In this work, by counting the adjacency matrices, we derive polynomial expressions that enumerate all connected, undirected, and unlabeled multigraphs with three nodes and fixed degree, and provide a method to efficiently generate them.

\end{abstract}

\maketitle

\section{Introduction}


Unlabeled multigraphs (such as those shown in Fig.~\ref{fig:defs}) find applications across various fields. In transportation, they are used to analyze the efficiency and structural properties of public transport networks~\cite{wang2020}. In social network analysis, they help study clustering, density, and recurring motifs~\cite{robins2013}. In data mining, unlabeled multigraphs are increasingly important for discovering frequent and meaningful patterns within complex, multi-relational data~\cite{ingalalli2018}. Additionally, they play a crucial role in robotics, particularly in multi-robot motion planning~\cite{vaisi2022}.

In polymer science, unlabeled multigraphs arise naturally in the statistical mechanics of polymer networks~\cite{Duplantier1986,Duplantier1989}, where the configurational entropy of the network depends critically on the underlying graph topology. This relationship is significant not only for materials science, since many polymeric materials are made of polymer networks~\cite{gu2020} or represented as multigraphs~\cite{Bonato2022,Bonato2025}, but also in biology, where identifying favored chromatin network structures is essential for understanding DNA transcription~\cite{Alberts2014,Brackley2013,Brackley2016,Cook2018,Chiang2025}—copying DNA into RNA~\cite{Calladine1997}—and organization.

Specifically, the enumeration and characterization of connected, traversable graphs are crucial because chromatin networks typically form when a chromatin fiber folds into a network through interactions with multivalent proteins~\cite{Brackley2013,Brackley2016}. By design, this network corresponds to a connected and traversable multigraph.

In this contest, to theoretically pin down preferred substructures or motifs, it is necessary to identify all unlabeled graphs and determine their statistical weights, which depend on both the number of distinct paths traversing each graph and the corresponding configurational entropy. While the BEST theorem~\cite{vA-deB1951,Fred1982,Stan1999} can be applied to calculate the number of paths in any multigraph~\cite{Bonato2025preprint}, estimating configurational entropy is generally feasible only under the approximation of phantom ideal chains~\cite{Bonato2024pre} or, when considering excluded volume effects, for asymptotically long polymers~\cite{Duplantier1989}.

In the latter case, the entropic scaling exponent regulating the configurational entropy is determined by the degree of the underlying unlabeled multigraph nodes. Therefore, identifying and enumerating all undirected, connected, and traversable unlabeled multigraphs with fixed node degrees is fundamental for predicting biologically relevant motifs in chromatin networks.

While Pólya’s theorem~\cite{redfield1927,polya1937,harary2018}, a classic and elegant application of group theory, provides generating functions for counting unlabeled multigraphs, manual calculations quickly become impractical as the number of nodes and edges grow, especially when node degree constraints are imposed. Theoretical solutions may be sought using adaptations of Pólya’s theorem, such as by Harary and Palmer~\cite{harary2018}, but their utilisation necessitates computational approaches even for relatively small graphs.

In the contest of analytical enumeration, our previous studies~\cite{Bonato2024,Bonato2024pre} have primarily focused on chromatin networks associated with two-node graphs. This work aims to extend the enumeration to unlabeled multigraphs with three nodes, focusing on traversable and connected graphs with fixed numbers of edges and node degrees.

The work is organised as follows. In Section II, we define the concepts of connected unlabeled traversable multigraphs. In Sections III and IV, we discuss some mathematical preliminaries, as the framework we use requires to partition and order the nodes in a given multigraph. We then provide the formulas and examples for the number of multigraphs for the case in which the three nodes have all different degrees (Section V, Eq.~\eqref{eq:counts_summed}), for that in which two have the same degree (Section VI, Eqs.~\eqref{eq:counts_summed_2} and ~\eqref{eq:counts_summed_3}), and for that in which they all have the same degree (Section VII, Eq.~\eqref{eq:counts_summed_4}). Finally, Section VIII contains our conclusions with an outlook for open problems which could be studied in the future.



\section{Connected unlabeled traversable multigraphs}

\begin{figure}
    \centering
    \includegraphics[width=0.65\linewidth]{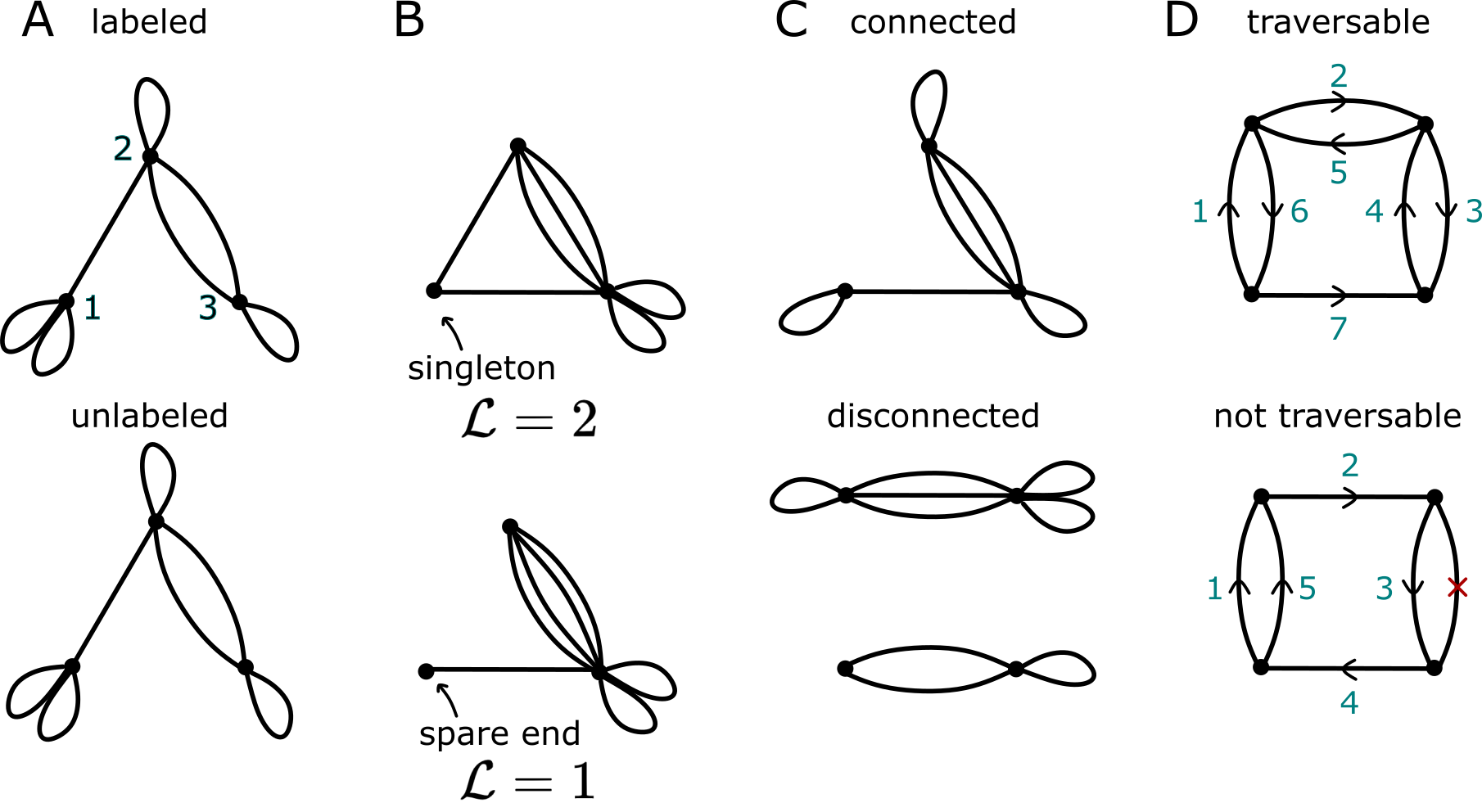}
    \caption{Examples of \textbf{A} labeled (top) and unlabeled (bottom) (multi)graphs, \textbf{B} graphs with a singleton ($\mathcal L = 2$, top) or an end ($\mathcal L = 1$, bottom), \textbf{C} connected (top) and disconnected (bottom) graphs, \textbf{D} traversable (top) and not traversable (bottom) graphs.}
    \label{fig:defs}
\end{figure}

We start with some definitions. 
\begin{definition}
A graph is an ordered pair $G=(V,E)$, where $V$ is a set of nodes (vertices), and $E$ a set of edges that connects them.
(Fig.~\ref{fig:defs}A, top)
\label{def:graph}
\end{definition}
In some contexts, graphs that allow multiple edges between two nodes or self-loops (edges connecting a node to itself) are called multigraphs. In this work, we focus on the more general multigraphs and do not distinguish between these variants (i.e. graph = multigraph).
Let us consider multigraphs with $V$ nodes.
We will denote $\mathcal L_i$ the degree of node $i$, namely the number of endpoints connected to node $i$ — we use endpoints and not edges, because an edge connecting node $i$ to itself contributes $+2$ to $\mathcal L_i$.
If $N$ is the total number of edges of a graph $G$, then 
\begin{equation}
2N = \sum_{i=1}^{V} \mathcal L_i.
\label{eq:deg_sum}
\end{equation}
If $\mathcal L_i \leq 2$, then node $i$ is a singleton or a spare end (Fig.~\ref{fig:defs}B).

\begin{figure}
    \centering
    \includegraphics[width=0.55\linewidth]{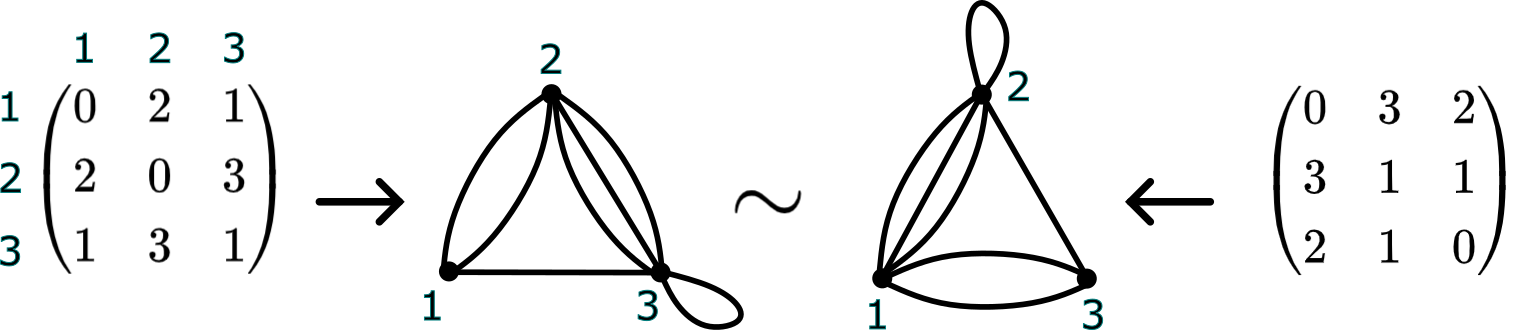}
    \caption{Example of mapping from adjacency matrices to multigraphs, and of isomorphic multigraphs and permutation-similar adjacency matrices. Each line (or column) of the matrix corresponds to a node of the graph, and the entries of that line to how that node is connected to the other nodes (i.e. number of connecting edges).}
    \label{fig:equi}
\end{figure}

We are interested in enumerating unlabeled connected and traversable graphs.
\begin{definition}
Two graphs are isomorphic if they are structurally identical. A graph isomorphism is a bijection between the vertex sets of the two graphs that preserves all their connections (Fig.~\ref{fig:equi}).
\label{def:isomorphic}
\end{definition}
\begin{definition}
Unlabeled graphs are isomorphism classes of graphs (Fig.~\ref{fig:defs}C).
\label{def:isomorphic}
\end{definition}
\begin{definition}
A graph is connected if there is at least one path from any node to any other node in the graph (Fig.~\ref{fig:defs}A).
\label{def:connected}
\end{definition}
\begin{definition}
A graph is traversable if there exists a path that traverses all edges of the graph without repetitions (Fig.~\ref{fig:defs}D).
\label{def:traversable}

\end{definition}
\begin{theorem}(\cite{Trudeau2013})
A disconnected graph is clearly not traversable, and it is well known that a connected graph is traversable if and only if exactly 0 or 2 of its nodes are of odd degree.
\label{th:trav_degrees}
\end{theorem}

We aim to establish a 1-to-1 correspondence between these graphs and a set of $V \times V$ matrices. We can then count the graphs by enumerating the matrices.

Given $ G_{V}$, the set of all unlabeled multigraphs with $V$ nodes, and $S_V$, the set of all symmetric $V \times V$ matrices, we can define, as shown in Fig.~\ref{fig:equi}, a map $g:\mathcal S_V \to G_V$ by associating each line of a matrix $M \in S_V$ with a node of a graph, and $M_{ij}$ with the number of edges joining nodes $i$ and $j$.
$M$ is the adjacency matrix of the graph.
Although this map is surjective, it is not injective, as different matrices can be mapped to isomorphic graphs (see Fig.~\ref{fig:equi}). In addition, $S_V$ contains adjacency matrices of disconnected (or, \textit{a priori} not traversable) graphs, which we want to exclude from the counting. Therefore, to properly enumerate the non-isomorphic graphs, we have to restrict the domain of $g$.
Consider an unlabeled graph $\mathcal G$.
To build its adjacency matrix, we have to label its nodes. Different labeling may produce different adjacency matrices.

\begin{definition}
Two adjacency matrices $A$ and $B$ are permutation-similar if one can be obtained from the other by simultaneous permutations of rows and columns, namely, if there exists a permutation matrix $P$ such that 
\begin{equation*}
PAP^{T} = B.
\end{equation*}
\end{definition}

\begin{theorem}(\cite{Trudeau2013})
Two multigraphs $\mathcal G_1$ and $\mathcal G_2$ are isomorphic if and only if their adjacency matrices are permutation-similar.
\label{th:theo_equivalence}
\end{theorem}

\section{Counting multigraphs with three nodes}

From now on, we set $V=3$.
As mentioned above, we are interested in connected and traversable graphs. Furthermore, inspired by the statistical mechanics of polymer networks, we are interested in the case $\mathcal L_i \geq 3$ $\forall i$, namely graphs without singletons and free ends (see Fig.~\ref{fig:defs}). A monodisperse polymer network with singletons is entropically equivalent to a polydisperse network without singletons~\cite{Duplantier1989}, whereas the enumeration of graphs with three nodes of which either one or two are free ends can be derived from our previous work~\cite{Bonato2024, Bonato2024pre}.

\subsection{Traversability}
Notice first that all connected graphs with three nodes are traversable.
Eq. \eqref{eq:deg_sum} states that the sum of the degrees of the nodes, $\mathcal L_1+\mathcal L_2+\mathcal L_3$, must be even; therefore, either two or none of $\mathcal L_1$, $\mathcal L_2$, and $\mathcal L_3$ are odd, and Theorem $\ref{th:trav_degrees}$ proves our statement.

\subsection{Connectivity}
Consider a symmetric $3\times 3$ matrix $M=\begin{pmatrix}
a & d & e \\
d & b & f \\
e & f & c
\end{pmatrix}\;$. $M$ is the adjacency matrix of a connected graph if and only if 
\begin{equation}
\begin{split}
d+e &\geq 1 \\
d+f &\geq 1 \\
e+f &\geq 1\;.
\end{split}
\label{eq:conn_eqs_3}
\end{equation}

\noindent Considering that 
\begin{equation}
\begin{split}
\mathcal L_1 &= 2a+d+e \\
\mathcal L_2 &= 2b+d+f \\
\mathcal L_3 &= 2c+e+f\,,
\end{split}
\label{eq:deg_eqs_3}\;
\end{equation}

\noindent Eq.~\ref{eq:conn_eqs_3} is equivalent to 

\begin{equation}
\begin{split}
\mathcal L_1-2a &\geq 1\,, \\
\mathcal L_2-2b &\geq 1\,, \\
\mathcal L_3-2c &\geq 1\;.
\end{split}
\label{eq:conn_deg_eqs_3}
\end{equation}

The interpretation of Eq.~\eqref{eq:conn_eqs_3} is that, for $V=3$, a graph is connected if, for every node there is at least one edge connecting it to another node. This is only true for $V=2,3$ as in this case, contrary to $V\geq 4$, there can be only one connected component with two or more nodes in a graph.

\section{Ordering the nodes}
To count only one adjacency matrix for every isomorphic class of graphs, we can exploit Theorem~\ref{th:theo_equivalence}.
Given a matrix $M$, we want to exclude from the counting all permutationally similar matrices. 
To factor these out, we assign an order to the nodes of the graph, which corresponds to inequalities in the coefficients of $M$.

We require $\mathcal L_1 \leq \mathcal L_2 \leq \mathcal L_3$.
If $\mathcal L_1 \neq \mathcal L_2 \neq \mathcal L_3$, this is sufficient to fix the order of the nodes; if two or all three degrees are equal, we must add additional prescriptions.

Noted that to draw a graph with $\mathcal L_i \geq3$, $i=1,2,3$ we need at least $N\geq5$ edges, the degrees of all possible graphs (without singletons and spare ends), in ascending order, obey Eq.~\eqref{eq:deg_sum} with $3\leq\mathcal L_1 \leq \mathcal L_2 \leq \mathcal L_3 $,
which is satisfied by $|3\leq\mathcal L_1\leq\mathcal L_2\leq\mathcal L_3|(N)$ combinations. If we denote the floor and ceiling functions with $\lfloor .\rfloor$ and $\lceil .\rceil$, since $\mathcal L_1+\mathcal L_2 + \mathcal L_3 = 2N$ and $\mathcal L_2 + \mathcal L_3\geq2\mathcal L_1$, then $3\leq\mathcal L_1\leq \left\lfloor \frac{2N}{3} \right\rfloor$, $\mathcal L_1\leq\mathcal L_2\leq \left\lfloor \frac{2N-\mathcal L_1}{2} \right\rfloor$ and $\mathcal L_3 = 2N - \mathcal L_1-\mathcal L_3$. Therefore,

\begin{equation}
|3\leq\mathcal L_1\leq\mathcal L_2\leq\mathcal L_3|(N) = \sum_{\mathcal L_1=3}^{\left\lfloor \frac{2N}{3} \right\rfloor}\sum_{\mathcal L_2=\mathcal L_1}^{\left\lfloor \frac{2N-\mathcal L_1 }{2} \right\rfloor} 1 =
(N-5)(\alpha-1)-\frac{3}{2}(\alpha-2)(\alpha-1)+(N-4)\beta -\frac{3}{2}(\beta-1)\beta\qquad N\geq5\;,
\label{eq:num_deg}
\end{equation}
\noindent where $\alpha= \left\lfloor\frac{\left\lfloor\frac{2N}{3}\right\rfloor}{2}\right\rfloor$ and $\beta= \left\lfloor\frac{\left\lfloor\frac{2N}{3} \right\rfloor-1}{2}\right\rfloor$. In Eq.~\ref{eq:num_deg}, $\mathcal L_1$ runs from 3 to $\left\lfloor\frac{2N}{3}\right\rfloor$ because 


The total number of and connected unlabeled graphs with $V=3$ and degrees $\geq 3$ will then be

\begin{equation}
\left. \left|G_3\right| = \sum_{\mathcal L_1=3}^{\left\lfloor \frac{2N}{3} \right\rfloor}\sum_{\mathcal L_2=\mathcal L_1}^{\left\lfloor \frac{2N-\mathcal L_1 }{2} \right\rfloor} \left|G_3(\mathcal L_1, \mathcal L_2, \mathcal L_3)\right|\,\right \rvert_{\mathcal L_3 = 2N-\mathcal L_1-\mathcal L_2}\;,
\label{eq:tot_counts_3}
\end{equation}
where $|G_3(\mathcal L_1, \mathcal L_2, \mathcal L_3)|$ is the number of graphs for fixed values of $\mathcal L_1$, $\mathcal L_2$ and $\mathcal L_3$.

In the following section, we will find $|G_3(\mathcal L_1, \mathcal L_2, \mathcal L_3)|$. We will distinguish between three cases: $\mathcal L_1 \neq \mathcal L_2 \neq \mathcal L_3$,
$\mathcal L_1 = \mathcal L_2 \neq \mathcal L_3$ or $\mathcal L_1 \neq \mathcal L_2 = \mathcal L_3$ and $\mathcal L_1 = \mathcal L_2 = \mathcal L_3$.
Note that
\begin{equation}
|3\leq\mathcal L_1=\mathcal L_2=\mathcal L_3|(N) = 
\begin{cases}
1 &\text{if}\;3|N\\
0 &\text{otherwise}
\end{cases}\qquad N\geq5\;,
\label{eq:num_deg_3eq}
\end{equation}

\begin{equation}
|3\leq\mathcal L_1<\mathcal L_2=\mathcal L_3|(N) = 
\sum_{p=2}^{\left\lfloor\frac{\left\lfloor \frac{2N}{3}\right\rfloor}{2}\right\rfloor}1-\begin{cases}
1 &\text{if}\;3|N\\
0 &\text{otherwise}
\end{cases}=\alpha-1-
\begin{cases}
1 &\text{if}\;3|N\\
0 &\text{otherwise}
\end{cases}\qquad N\geq5\;,
\label{eq:num_deg_2eq_1}
\end{equation}

\begin{equation}
|3\leq\mathcal L_1=\mathcal L_2<\mathcal L_3|(N) = 
\sum_{p=\left\lceil\frac{\left\lceil \frac{2N}{3}\right\rceil}{2}\right\rceil}^{N-3}1-\begin{cases}
1 &\text{if}\;3|N\\
0 &\text{otherwise}
\end{cases}=N-2-\gamma-
\begin{cases}
1 &\text{if}\;3|N\\
0 &\text{otherwise}
\end{cases}\qquad N\geq5\;,
\label{eq:num_deg_2eq_2}
\end{equation}
where $\gamma = \left\lceil\frac{\left\lceil \frac{2N}{3}\right\rceil}{2}\right\rceil$, $3|N$ means $N$ is a multiple of 3, and

\begin{equation}
\begin{split}
|3\leq\mathcal L_1<\mathcal L_2<\mathcal L_3|(N) &= |3\leq\mathcal L_1\leq\mathcal L_2\leq\mathcal L_3|(N)- |3\leq\mathcal L_1=\mathcal L_2<\mathcal L_3|(N)\\&-|3\leq\mathcal L_1<\mathcal L_2=\mathcal L_3|(N)-|3\leq\mathcal L_1=\mathcal L_2=\mathcal L_3|(N)\\
= (N-5)(\alpha-1)&-\frac{3}{2}(\alpha-2)(\alpha-1)+(N-4)\beta -\frac{3}{2}(\beta-1)\beta-\alpha+\gamma-N+3 +\begin{cases}
1 &\text{if}\;3|N\\
0 &\text{otherwise}
\end{cases}
\qquad N\geq5\;.
\end{split}
\label{eq:num_deg_3diff}
\end{equation}

\section{Different degrees}

As discussed above, if $\mathcal L_1 \neq \mathcal L_2 \neq \mathcal L_3$, then sorting the degrees in ascending order is enough to factor the relabeling of the nodes out.

Since Eq.~\eqref{eq:deg_eqs_3} imposes three constraints on the six free coefficients of $M$, we can express $d$, $e$ and $f$ as functions of $a$, $b$ and $c$:

\begin{equation}
\begin{split}
\frac{\mathcal L_1 + \mathcal L_2- \mathcal L_3-2a-2b+2c}{2} &= d \\
\frac{\mathcal L_1 - \mathcal L_2+ \mathcal L_3-2a+2b-2c}{2} &= e \\
\frac{-\mathcal L_1 + \mathcal L_2+ \mathcal L_3+2a-2b-2c}{2} &= f\;.
\end{split}
\label{eq:def_3}
\end{equation}

To count all non-isomorphic configurations, we have to sum over all possible values of $a$, $b$ and $c$ compatible with Eq.~\eqref{eq:conn_deg_eqs_3}, which ensures connectivity, and then remove solutions that satisfy one of these three equations:

\begin{equation}
\begin{split}
\frac{\mathcal L_1 + \mathcal L_2- \mathcal L_3-2a-2b+2c}{2} &< 0 \\
\frac{\mathcal L_1 - \mathcal L_2+ \mathcal L_3-2a+2b-2c}{2} &< 0 \\
\frac{-\mathcal L_1 + \mathcal L_2+ \mathcal L_3+2a-2b-2c}{2} &< 0\;.
\end{split}
\label{eq:constraint_3}
\end{equation}

\textit{A priori}, the sets of solutions satisfying each of these three equations could have non-empty intersection, and considering them individually could lead to overcounting the matrices we have to remove, but if we impose Eq.~\eqref{eq:conn_deg_eqs_3}, which is equivalent to Eq.~\eqref{eq:conn_eqs_3}, then there cannot be a solution $a$, $b$ and $c$ satisfying more than one of these equations simultaneously.

Finally, the number of unlabeled connected (and traversable) graphs with degrees $\mathcal L_1$, $\mathcal L_2$, $\mathcal L_3$, and $\mathcal L_1 \neq \mathcal L_2 \neq \mathcal L_3$ is 

\begin{equation}
\begin{split}
|G(3\leq\mathcal L_1<\mathcal L_2<\mathcal L_3)| = \sum_{a=0}^{\left\lfloor \frac{\mathcal L_1-1}{2} \right\rfloor}\sum_{ b=0}^{\left\lfloor \frac{\mathcal L_2-1}{2} \right\rfloor}\sum_{ c=0}^{\left\lfloor \frac{\mathcal L_3-1}{2} \right\rfloor} 1&-\sum_{ a=0}^{\left\lfloor \frac{\mathcal L_1-1}{2} \right\rfloor}\sum_{ b=0}^{\left\lfloor \frac{\mathcal L_2-1}{2} \right\rfloor}\sum_{ c=0}^{\min\left(\left\lfloor \frac{\mathcal L_3-1}{2} \right\rfloor,a+b-\frac{\mathcal L_1+\mathcal L_2-\mathcal L_3}{2}\right)} 1 \\
&-\sum_{ a=0}^{\left\lfloor \frac{\mathcal L_1-1}{2} \right\rfloor}\sum_{ b=0}^{\left\lfloor \frac{\mathcal L_2-1}{2} \right\rfloor}\sum_{ c=\max\left(0,-a+b+\frac{\mathcal L_1-\mathcal L_2+\mathcal L_3}{2}+1\right)}^{\left\lfloor \frac{\mathcal L_3-1}{2} \right\rfloor} 1 \\
&-\sum_{ a=0}^{\left\lfloor \frac{\mathcal L_1-1}{2} \right\rfloor}\sum_{ b=0}^{\left\lfloor \frac{\mathcal L_2-1}{2} \right\rfloor}\sum_{ c=\max\left(0,a-b+\frac{-\mathcal L_1+\mathcal L_2+\mathcal L_3}{2}+1\right)}^{\left\lfloor \frac{\mathcal L_3-1}{2} \right\rfloor} 1\; ,
\label{eq:number_ineq_1_3}
\end{split}
\end{equation}
\noindent where the second to forth terms count the solutions to Eq. \eqref{eq:constraint_3}. Note that the upper bounds of the sums account for Eq.~\eqref{eq:conn_deg_eqs_3}  (and Eq.~\eqref{eq:conn_eqs_3}).

As detailed in Appendix~\ref{appendix:A}, Eq.~\eqref{eq:number_ineq_1_3} can be explicitly summed to obtain the polynomial expression

\begin{equation}
\begin{split}
|G(3\leq\mathcal L_1<\mathcal L_2<\mathcal L_3)|&=\prod_{i=1}^3 \left\lfloor \frac{\mathcal L_i+1}{2} \right\rfloor-
\begin{cases}
g_1(A_1,B_1,K_1)&\text{if}\;\mathcal L_1+\mathcal L_2+ 1\geq\mathcal L_3 \geq \mathcal L_2+3\\
g_2(A_1,B_1,K_1)&\text{if}\;\mathcal L_1+\mathcal L_2+ 1\leq\mathcal L_3\\
0 &\text{otherwise}
\end{cases}\\
&-\begin{cases}
g_3(A_1,B_1,K_1)&\text{if}\;
\left(\mathcal L_2 \geq \mathcal L_1+2 \land 2|(\mathcal L_1+1)\right)\lor(2|\mathcal L_1)\\ 
0&\text{otherwise}
\end{cases}
-g_4(A_1,B_1,K_1)\\
&-\begin{cases}
g_5(A_2,B_2,K_2)&\text{if}\;\mathcal L_2 \geq \mathcal L_1+3\\
0&\text{otherwise}
\end{cases}
-g_4(A_2,B_2,K_2)-g_6(A_3,B_3,K_3)\,,
\end{split}
\label{eq:counts_summed}
\end{equation}
where $\land $ and $\lor$ denote logical AND and OR, and

\begin{equation}
\begin{split}
g_1(A,B,K):=&(A-K)\left(K+2+\frac{(A-K-1)(K+2A+8)}{6}\right) \\
g_2(A,B,K):=&(A+1)\left(-K+\frac{A(2A+4-3K)}{6}\right) \\
g_3(A,B,K):=&\frac{(A+1)(B-A)(A+B-2K+1)}{2} \\
g_4(A,B,K):=&A\left(A(B-K+1)+\frac{(A-1)(A-3B+3K-2)}{6}\right) \\
g_5(A,B,K):=&\frac{(A+1)(B-K)(B-K+1)}{2} \\
g_6(A,B,K):=&(A+B-K)\left(A+B-K+\frac{(A+B-K-1)(A+B-K-2)}{6}\right)\:, \\
\end{split}
\label{eq:gs}
\end{equation}

and $A_i$, $B_i$, $K_i$ and $\mathcal L_i$ are defined in Table~\ref{tab:coeffs}.

\renewcommand{\arraystretch}{1.7} 
\begin{table}[h!]
\centering
\begin{tabular}{|c|c|c|c|c|}
\hline
$i$ & $A_i$ & $B_i$ & $C_i$ & $K_i$ \\
\hline
$1$ & $\left\lfloor \frac{\mathcal L_1-1}{2} \right\rfloor$ & $\left\lfloor \frac{\mathcal L_2-1}{2} \right\rfloor$ & $\left\lfloor \frac{\mathcal L_3-1}{2} \right\rfloor$ & $\frac{\mathcal L_1+\mathcal L_2-\mathcal L_3}{2}$ \\
$2$ & $\left\lfloor \frac{\mathcal L_1-1}{2} \right\rfloor$ & $\left\lfloor \frac{\mathcal L_3-1}{2} \right\rfloor$ & $\left\lfloor \frac{\mathcal L_2-1}{2} \right\rfloor$ & $\frac{\mathcal L_1-\mathcal L_2+\mathcal L_3}{2}$ \\
$3$ & $\left\lfloor \frac{\mathcal L_2-1}{2} \right\rfloor$ & $\left\lfloor \frac{\mathcal L_3-1}{2} \right\rfloor$ & $\left\lfloor \frac{\mathcal L_1-1}{2} \right\rfloor$ & $\frac{-\mathcal L_1+\mathcal L_2+\mathcal L_3}{2}$ \\
\hline
\end{tabular}
\caption{Coefficients in Eq.~\eqref{eq:counts_summed}.}
\label{tab:coeffs}
\end{table}


Eqs.~\eqref{eq:def_3} and~\eqref{eq:constraint_3} can also be used to systematically label and draw all unlabeled graphs: given all combinations of $a< \lfloor\frac{\mathcal L_1-1}{2}\rfloor$, $b< \left\lfloor\frac{\mathcal L_2-1}{2}\right\rfloor$ and $c< \left\lfloor\frac{\mathcal L_3-1}{2}\right\rfloor$, we can compute $d$, $e$ and $f$ with Eq.~\eqref{eq:def_3}, discard invalid solutions with Eq.~\eqref{eq:constraint_3} (namely these $a$, $b$ and $c$ which yeld negative $d$, $e$ or $f$) and then draw the graphs from the adjacency matrices. We next give an example.

\subsection{Example with different degrees}
\label{sec:example_diff}

\begin{figure}[t!]
    \centering
    \includegraphics[width=0.85\linewidth]{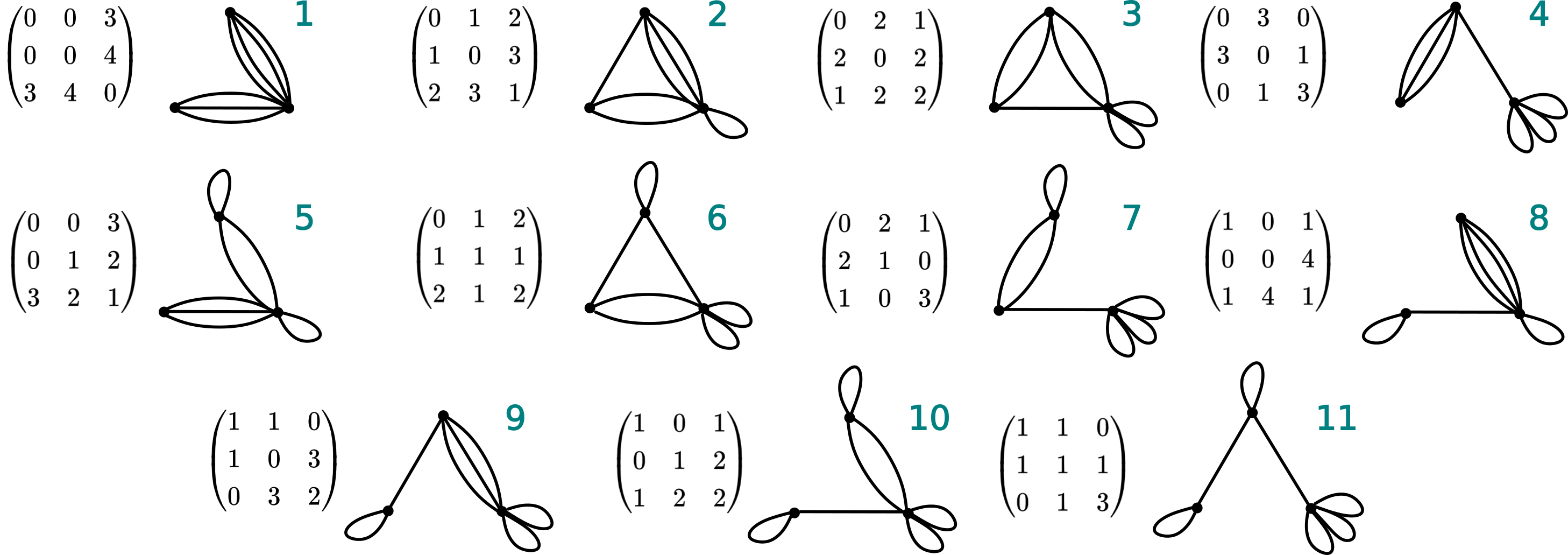}
    \caption{Connected unlabeled graphs with $\mathcal L_1=3$, $\mathcal L_2=4$ and $\mathcal L_3=7$.}
    \label{fig:ineq347}
\end{figure}

\begin{figure}
    \centering
    \includegraphics[width=0.95\linewidth]{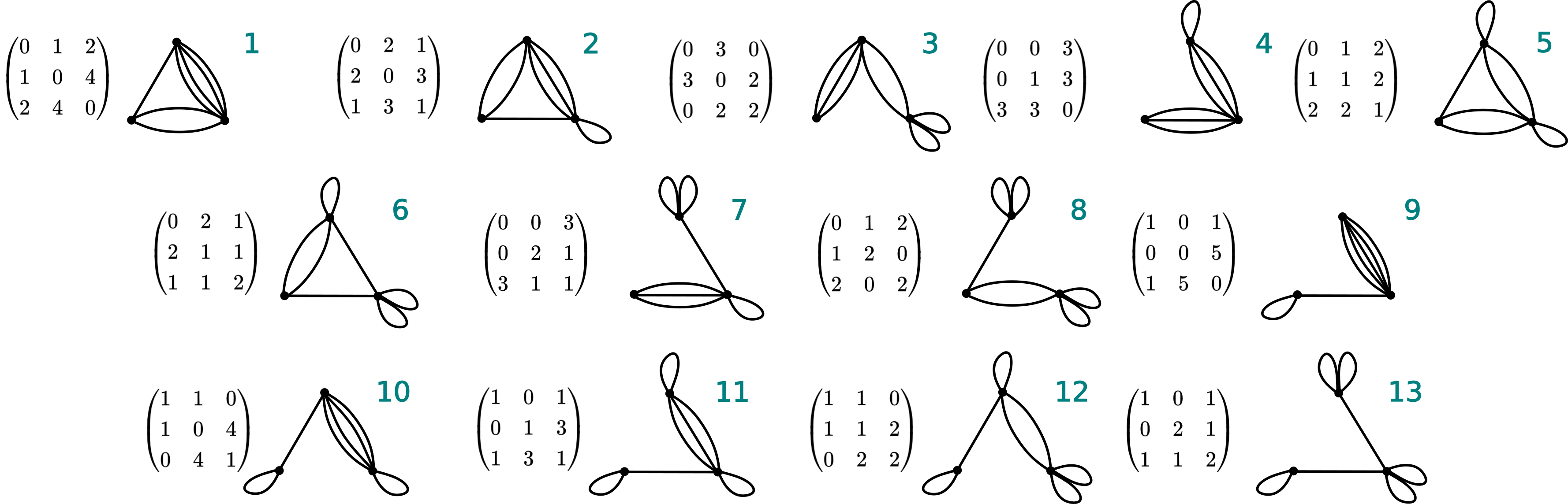}
    \caption{Connected unlabeled graphs with $\mathcal L_1=3$, $\mathcal L_2=5$ and $\mathcal L_3=6$.}
    \label{fig:ineq356}
\end{figure}

Since the number of graphs $|G(\mathcal L_1, \mathcal L_2, \mathcal L_3)|$ increases rapidly with the number of edges $N$, we will draw all unlabeled graphs for $N=7$, which was the focus of previous work~\cite{Bonato2024,Bonato2024pre}, and report only $|G(\mathcal L_1, \mathcal L_2, \mathcal L_3)|$ for larger values of $N$.
Our results were tested by brute force enumeration, namely by computationally generating all symmetric matrices $M$ satisfying $\sum_jM_{ij}=\mathcal L_i$, and all their permutations, and then using Theorem~\ref{th:theo_equivalence}.
As enumerated by Eq.~\eqref{eq:num_deg_3diff}, there are only two possible combinations $3\leq\mathcal L_1 <\mathcal L_2 < \mathcal L_3$ that satisfy $\mathcal L_1+\mathcal L_2+\mathcal L_3 = 2N=14$, these are $\mathcal L_1,\mathcal L_2, \mathcal L_3=3,4,7$ and $\mathcal L_1,\mathcal L_2, \mathcal L_3=3,5,6$.
Eq.~\eqref{eq:counts_summed} gives $|G(3,4,7)|=11$ and $|G(3,5,6)|=13$. The corresponding adjacency matrices are obtained using Eqs.~\eqref{eq:def_3} and~\eqref{eq:constraint_3} as explained above. The entries of the adjacency matrices are reported in Table~\ref{tab:ineq_table_diff}, the corresponding graphs are drawn in Figs.~\ref{fig:ineq347} and~\ref{fig:ineq356}. In Tables~\ref{tab:ineq_table_6to10} and~\ref{tab:ineq_table_15} (Appendix \ref{appendix:B}) we report the number of unlabeled graphs with $N=6,7,8,9,10$, and $15$ edges, obtained with Eq.~\eqref{eq:counts_summed}.

\begin{table}[h!]
\centering
\begin{tabular}{|*{14}{c|}}
\hline
\multicolumn{7}{|c|}{$\mathcal L_1=3$, $\mathcal L_2=4$, $\mathcal L_3=7$} & \multicolumn{7}{c|}{$\mathcal L_1=3$, $\mathcal L_2=5$, $\mathcal L_3=6$} \\
\hline
index & $a$ & $b$ & $c$ & $d$ & $e$ & $f$ & index & $a$ & $b$ & $c$ & $d$ & $e$ & $f$\\
\hline
$\textcolor{blue}{1}$ & $0$ & $0$ & $0$ & $0$ & $3$ & $4$ & $\textcolor{blue}{1}$ & $0$ & $0$ & $0$ & $1$ & $2$ & $4$\\
$\textcolor{blue}{2}$ & $0$ & $0$ & $1$ & $1$ & $2$ & $3$& $\textcolor{blue}{2}$ & $0$ & $0$ & $1$ & $2$ & $1$ & $3$\\
$\textcolor{blue}{3}$ & $0$ & $0$ & $2$ & $2$ & $1$ & $2$ & $\textcolor{blue}{3}$ & $0$ & $0$ & $2$ & $3$ & $0$ & $2$\\
$\textcolor{blue}{4}$ & $0$ & $0$ & $3$ & $3$ & $0$ & $1$ & $\textcolor{blue}{4}$ & $0$ & $1$ & $0$ & $0$ & $3$ & $3$\\
$\textcolor{red}{1i}$ & $\textcolor{red}{0}$ & $\textcolor{red}{1}$ & $\textcolor{red}{0}$ & $\textcolor{red}{-1}$ & $\textcolor{red}{4}$ & $\textcolor{red}{3}$ & $\textcolor{blue}{5}$ & $0$ & $1$ & $1$ & $1$ & $2$ & $2$\\
$\textcolor{blue}{5}$ & $0$ & $1$ & $1$ & $0$ & $3$ & $2$ & $\textcolor{blue}{6}$ & $0$ & $1$ & $2$ & $2$ & $1$ & $1$\\
$\textcolor{blue}{6}$ & $0$ & $1$ & $2$ & $1$ & $2$ & $1$ & $\textcolor{red}{1i}$ & $\textcolor{red}{0}$ & $\textcolor{red}{2}$ & $\textcolor{red}{0}$ & $\textcolor{red}{-1}$ & $\textcolor{red}{4}$ & $\textcolor{red}{2}$ \\
$\textcolor{blue}{7}$ & $0$ & $1$ & $3$ & $2$ & $1$ & $0$ & $\textcolor{blue}{7}$ & $0$ & $2$ & $1$ & $0$ & $3$ & $1$\\
$\textcolor{red}{2i}$ & $\textcolor{red}{1}$ & $\textcolor{red}{0}$ & $\textcolor{red}{0}$ & $\textcolor{red}{-1}$ & $\textcolor{red}{2}$ & $\textcolor{red}{5}$ & $\textcolor{blue}{8}$ & $0$ & $2$ & $2$ & $1$ & $2$ & $0$\\
$\textcolor{blue}{8}$ & $1$ & $0$ & $1$ & $0$ & $1$ & $4$ & $\textcolor{blue}{9}$ & $1$ & $0$ & $0$ & $0$ & $1$ & $5$\\
$\textcolor{blue}{9}$ & $1$ & $0$ & $2$ & $1$ & $0$ & $3$ & $\textcolor{blue}{10}$ & $1$ & $0$ & $1$ & $1$ & $0$ & $4$\\
$\textcolor{red}{3i}$ & $\textcolor{red}{1}$ & $\textcolor{red}{0}$ & $\textcolor{red}{3}$ & $\textcolor{red}{2}$ & $\textcolor{red}{-1}$ & $\textcolor{red}{2}$ & $\textcolor{red}{2i}$ & $\textcolor{red}{1}$ & $\textcolor{red}{0}$ & $\textcolor{red}{2}$ & $\textcolor{red}{2}$ & $\textcolor{red}{-1}$ & $\textcolor{red}{3}$\\
$\textcolor{red}{4i}$ & $\textcolor{red}{1}$ & $\textcolor{red}{1}$ & $\textcolor{red}{0}$ & $\textcolor{red}{-2}$ & $\textcolor{red}{3}$ & $\textcolor{red}{4}$ & $\textcolor{red}{3i}$ & $\textcolor{red}{1}$ & $\textcolor{red}{1}$ & $\textcolor{red}{0}$ & $\textcolor{red}{-1}$ & $\textcolor{red}{2}$ & $\textcolor{red}{4}$\\
$\textcolor{red}{5i}$ & $\textcolor{red}{1}$ & $\textcolor{red}{1}$ & $\textcolor{red}{1}$ & $\textcolor{red}{-1}$ & $\textcolor{red}{2}$ & $\textcolor{red}{3}$ & $\textcolor{blue}{11}$ & $1$ & $1$ & $1$ & $0$ & $1$ & $3$\\
$\textcolor{blue}{10}$ & $1$ & $1$ & $2$ & $0$ & $1$ & $2$ & $\textcolor{blue}{12}$ & $1$ & $1$ & $2$ & $1$ & $0$ & $2$\\
$\textcolor{blue}{11}$ & $1$ & $1$ & $3$ & $1$ & $0$ & $1$ & $\textcolor{red}{4i}$ & $\textcolor{red}{1}$ & $\textcolor{red}{2}$ & $\textcolor{red}{0}$ & $\textcolor{red}{-1}$ & $\textcolor{red}{3}$ & $\textcolor{red}{3}$\\
& &  & &  & &  & $\textcolor{red}{5i}$ & $\textcolor{red}{1}$ & $\textcolor{red}{2}$ & $\textcolor{red}{1}$ & $\textcolor{red}{-1}$ & $\textcolor{red}{2}$ & $\textcolor{red}{2}$\\
& & & & & & & $\textcolor{blue}{13}$ & $1$ & $2$ & $2$ & $0$ & $1$ & $1$\\
\hline

\end{tabular}
\caption{Unlabeled graphs with $\mathcal L_1=3$, $\mathcal L_2=4$ and $\mathcal L_3=7$ (left) and $\mathcal L_1=3$, $\mathcal L_2=5$ and $\mathcal L_3=6$ (right). Entries of the adjacency matrices. Invalid topologies are colored in red.}
\label{tab:ineq_table_diff}
\end{table}

\section{Two equal degrees}

If $\mathcal L_2=\mathcal L_3$, then we need to order nodes 2 and 3 differently. This can be done by requiring $c\geq b$. This prescription allows us to count only once these graphs that are symmetric under the permutation of nodes 2 and 3 and have multiple permutation-similar adjacency matrices. Note that the only solution for $b=c$, $\mathcal L_2=\mathcal L_3$ is of the form
\renewcommand{\arraystretch}{1.0} 
\begin{equation*}
M = \begin{pmatrix}
a & d & d \\
d & b & f \\
d & f & b
\end{pmatrix}\; ,
\end{equation*}
\renewcommand{\arraystretch}{1.7} 
which is invariant under the permutation of nodes $(2,3)$. This means that adjacency matrices with $b=c$ and $\mathcal L_1<\mathcal L_2=\mathcal L_3$ are in 1 to 1 correspondence with non-isomorphic graphs; therefore, when $b=c$, we do not require any additional order constraints. The number of unlabeled graphs in this case is

\begin{equation}
\begin{split}
|G(3\leq\mathcal L_1<\mathcal L_2=\mathcal L_3)| = \sum_{a=0}^{\left\lfloor \frac{\mathcal L_1-1}{2} \right\rfloor}\sum_{ b=0}^{\left\lfloor \frac{\mathcal L_2-1}{2} \right\rfloor}\sum_{ c=b}^{\left\lfloor \frac{\mathcal L_3-1}{2} \right\rfloor} 1&-\sum_{ a=0}^{\left\lfloor \frac{\mathcal L_1-1}{2} \right\rfloor}\sum_{ b=0}^{\left\lfloor \frac{\mathcal L_2-1}{2} \right\rfloor}\sum_{ c=b}^{\min\left(\left\lfloor \frac{\mathcal L_3-1}{2} \right\rfloor,a+b-\frac{\mathcal L_1+\mathcal L_2-\mathcal L_3}{2}\right)} 1 \\
&-\sum_{ a=0}^{\left\lfloor \frac{\mathcal L_1-1}{2} \right\rfloor}\sum_{ b=0}^{\left\lfloor \frac{\mathcal L_2-1}{2} \right\rfloor}\sum_{ c=\max\left(b,-a+b+\frac{\mathcal L_1-\mathcal L_2+\mathcal L_3}{2}+1\right)}^{\left\lfloor \frac{\mathcal L_3-1}{2} \right\rfloor} 1 \\
&-\sum_{ a=0}^{\left\lfloor \frac{\mathcal L_1-1}{2} \right\rfloor}\sum_{ b=0}^{\left\lfloor \frac{\mathcal L_2-1}{2} \right\rfloor}\sum_{ c=\max\left(b,a-b+\frac{-\mathcal L_1+\mathcal L_2+\mathcal L_3}{2}+1\right)}^{\left\lfloor \frac{\mathcal L_3-1}{2} \right\rfloor} 1\; .
\label{eq:number_ineq_2_3}
\end{split}
\end{equation}

If we define

\begin{equation}
\begin{split}
g_7(A,B):=&\frac{1}{2}(A+1)\left((B-A)(B-A-1)+\frac{1}{3}A(2B-2A-1)\right) \\
g_8(A,B,p):=&\sum_{a=0}^A\frac{1}{2}(D_1(a,p)-B+A-a-p)(D_1(a,p)-B+A-a-p+1) \\
+&\sum_{a=0}^A\frac{1}{2}(B-D_2(a,p)+1)(B-D_2(a,p)+2)\;,
\\
\end{split}
\label{eq:gs_2}
\end{equation}
where $D_1(a,p)=\left\lfloor\frac{2B-A-1-p+a}{2}\right\rfloor$, and $D_2(a,p)=\left\lceil\frac{2B-A-1-p+a}{2}\right\rceil$, then

\begin{equation}
\begin{split}
|G(3\leq\mathcal L_1<\mathcal L_2=\mathcal L_3)|&=\frac{1}{2}\left\lfloor\frac{\mathcal L_1+1}{2}\right\rfloor\left\lfloor\frac{\mathcal L_2+1}{2}\right\rfloor\left\lfloor\frac{\mathcal L_2+3}{2}\right\rfloor-g_7(A_1,B_1)-g_8\left(A_1,B_1,
p=\begin{cases}
2 &2 |\mathcal L_2\\
1 & \text{otherwise}   
\end{cases}\right)
\end{split}\;.
\label{eq:counts_summed_2}
\end{equation}

\noindent Similarly, if $\mathcal L_1=\mathcal L_2$, 

\begin{equation}
\begin{split}
|G(3\leq\mathcal L_1=\mathcal L_2<\mathcal L_3)| = \sum_{a=0}^{\left\lfloor \frac{\mathcal L_1-1}{2} \right\rfloor}\sum_{ b=a}^{\left\lfloor \frac{\mathcal L_2-1}{2} \right\rfloor}\sum_{ c=0}^{\left\lfloor \frac{\mathcal L_3-1}{2} \right\rfloor} 1&-\sum_{ a=0}^{\left\lfloor \frac{\mathcal L_1-1}{2} \right\rfloor}\sum_{ b=a}^{\left\lfloor \frac{\mathcal L_2-1}{2} \right\rfloor}\sum_{ c=0}^{\min\left(\left\lfloor \frac{\mathcal L_3-1}{2} \right\rfloor,a+b-\frac{\mathcal L_1+\mathcal L_2-\mathcal L_3}{2}\right)} 1 \\
&-\sum_{ a=0}^{\left\lfloor \frac{\mathcal L_1-1}{2} \right\rfloor}\sum_{ b=a}^{\left\lfloor \frac{\mathcal L_2-1}{2} \right\rfloor}\sum_{ c=\max\left(0,-a+b+\frac{\mathcal L_1-\mathcal L_2+\mathcal L_3}{2}+1\right)}^{\left\lfloor \frac{\mathcal L_3-1}{2} \right\rfloor} 1 \\
&-\sum_{ a=0}^{\left\lfloor \frac{\mathcal L_1-1}{2} \right\rfloor}\sum_{ b=a}^{\left\lfloor \frac{\mathcal L_2-1}{2} \right\rfloor}\sum_{ c=\max\left(0,a-b+\frac{-\mathcal L_1+\mathcal L_2+\mathcal L_3}{2}+1\right)}^{\left\lfloor \frac{\mathcal L_3-1}{2} \right\rfloor} 1\; ,
\label{eq:number_ineq_3_3}
\end{split}
\end{equation}

\noindent and, if we define

\begin{equation}
\begin{split}
g_{9}(A,C,p):=&(A-T(p)+1)\left(\frac{1}{2}(A+1)(-3A-2p+2+2C)\right. \\
+&\left.\frac{1}{4}(A+T(p))(6A+1+2p-2C)-\frac{3}{2}(T(p)^2+(A-T(p))(4T(p)+2A+1)/6)\right)\\
+&\frac{1}{2}(D_3(p)+1)\left((C+1-A-p)(C+2-A-p)+D_3(p)\left(\frac{2C+3-2p-2A}{2}+\frac{2D_3(p)+1}{6}\right) \right),
\\
g_{10}(A):=&\frac{(A-1)A(A+1)}{6} \\
\end{split}
\label{eq:gs_3}
\end{equation}
where $D_3(p)=\left\lfloor\frac{2A+p-C-1}{2}\right\rfloor$, and $T(p)=\max(0,D_3(p)+1)$,
\begin{equation}
\begin{split}
|G(3\leq\mathcal L_1=\mathcal L_2<\mathcal L_3)|&=\frac{1}{2}\left\lfloor\frac{\mathcal L_1+1}{2}\right\rfloor\left\lfloor\frac{\mathcal L_1+3}{2}\right\rfloor\left\lfloor\frac{\mathcal L_3+1}{2}\right\rfloor-g_{9}\left(A_1,C_1,p=\begin{cases}
2 &\text{if}\;2 |\mathcal L_1\\
1 &\text{otherwise}   
\end{cases}\right)-g_{10}(A_1)
\end{split}\;.
\label{eq:counts_summed_3}
\end{equation}

\subsection{Example with two equal degrees}
\begin{figure}[h!]
    \centering
    \includegraphics[width=0.95\linewidth]{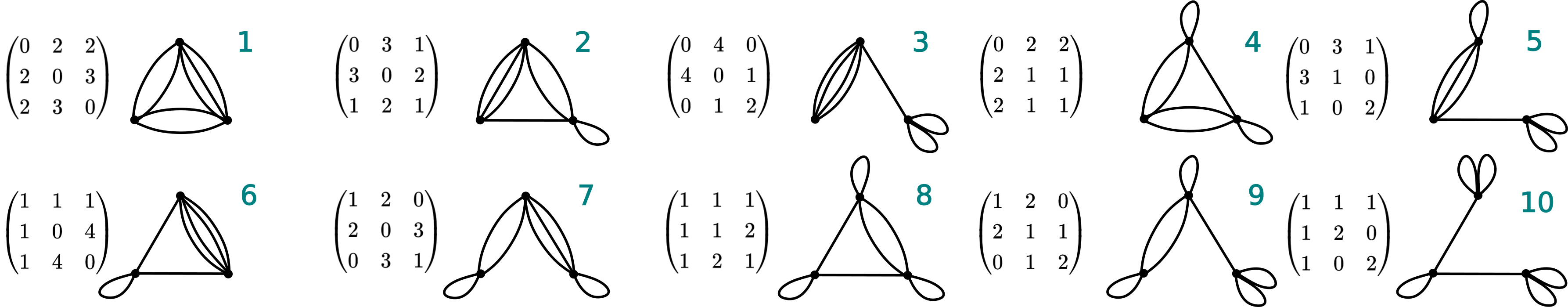}
    \caption{Connected unlabeled graphs with $\mathcal L_1=4$ and $\mathcal L_2 = \mathcal L_3=5$.}
    \label{fig:ineq455}
\end{figure}

\begin{figure}[h!]
    \centering
    \includegraphics[width=0.65\linewidth]{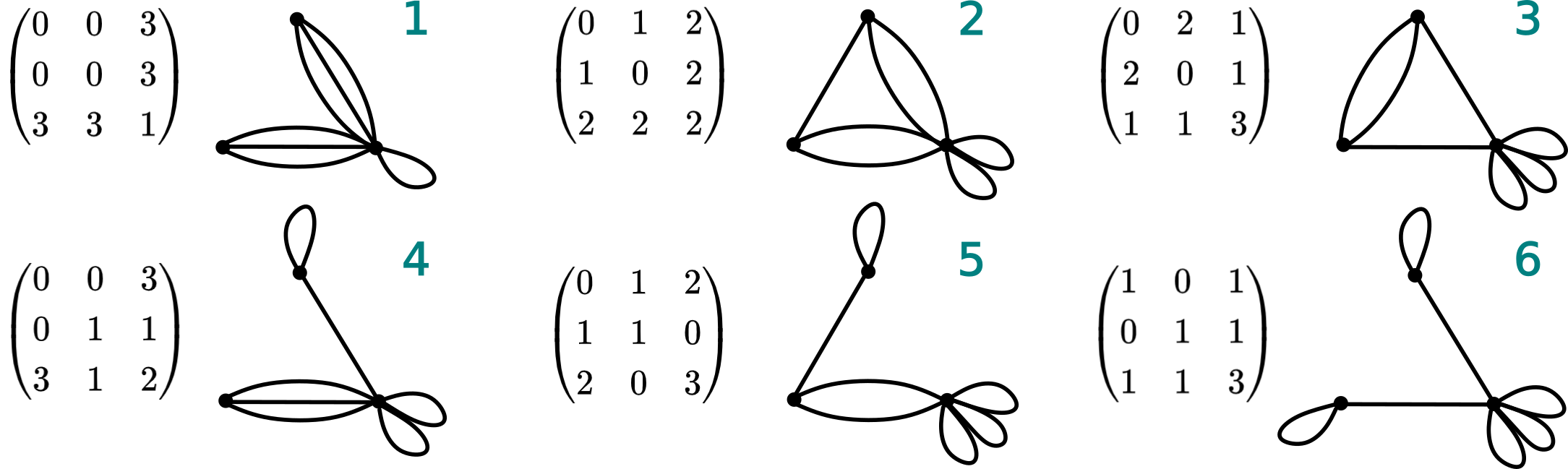}
    \caption{Connected unlabeled graphs with $\mathcal L_1=\mathcal L_2=3$ and $\mathcal L_3=8$.}
    \label{fig:ineq338}
\end{figure}

\begin{figure}[h!]
    \centering
    \includegraphics[width=0.85\linewidth]{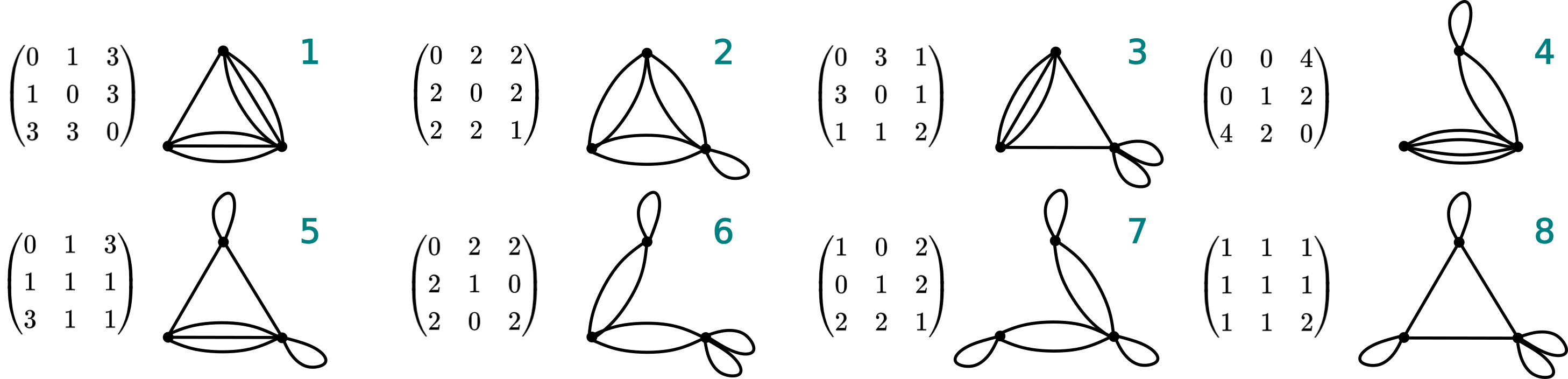}
    \caption{Connected unlabeled graphs with $\mathcal L_1=\mathcal L_2=4$ and $\mathcal L_3=6$.}
    \label{fig:ineq446}
\end{figure}

Here we complete the example reported in Section \ref{sec:example_diff} by counting the number of unlabeled graphs with $N=7$ and $3\leq\mathcal L_1 = \mathcal L_2 < \mathcal L_3$, or $3\leq\mathcal L_1 < \mathcal L_2 = \mathcal L_3$. Eqs. \eqref{eq:num_deg_2eq_1} and \eqref{eq:num_deg_2eq_2} tell us that $|3\leq\mathcal L_1<\mathcal L_2=\mathcal L_3|(7)=1$ and $|3\leq\mathcal L_1=\mathcal L_2<\mathcal L_3|(7)=2$. The corresponding combinations are $\mathcal L_1, \mathcal L_2, \mathcal L_3=4,5,5$; $3,3,8$ and $4,4,6$. Eqs. \eqref{eq:counts_summed_2} and \eqref{eq:counts_summed_3} then give $|G(4,5,5)|=10$, $|G(3,3,8)|=6$ and $|G(4,4,6)|=8$. Tables \ref{tab:ineq_table_L2eL3} and \ref{tab:ineq_table_L1eL2} report all unlabeled graphs for these three cases, graphs which are also drawn in Figs. \ref{fig:ineq455}, \ref{fig:ineq338} and \ref{fig:ineq446}. Finally, Table \ref{tab:ineq_table_6to15} (Appendix \ref{appendix:B}) collects the number of unlabeled graphs with $3\leq\mathcal L_1 = \mathcal L_2 < \mathcal L_3$, or $3\leq\mathcal L_1 < \mathcal L_2 = \mathcal L_3$ and $N=6,7,8,9,10$ and $15$.

\begin{table}[h!]
\centering
\begin{tabular}{|*{7}{c|}}
\hline
\multicolumn{7}{|c|}{$\mathcal L_1=4$, $\mathcal L_2=5$, $\mathcal L_3=5$}  \\
\hline
index & $a$ & $b$ & $c$ & $d$ & $e$ & $f$\\
\hline
$\textcolor{blue}{1}$ & $0$ & $0$ & $0$ & $2$ & $2$ & $3$\\
$\textcolor{blue}{2}$ & $0$ & $0$ & $1$ & $3$ & $1$ & $2$\\
$\textcolor{blue}{3}$ & $0$ & $0$ & $2$ & $4$ & $0$ & $1$\\
$\textcolor{blue}{4}$ & $0$ & $1$ & $1$ & $2$ & $2$ & $1$\\
$\textcolor{blue}{5}$ & $0$ & $1$ & $2$ & $3$ & $1$ & $0$\\
$\textcolor{red}{1i}$ & $\textcolor{red}{0}$ & $\textcolor{red}{2}$ & $\textcolor{red}{2}$ & $\textcolor{red}{2}$ & $\textcolor{red}{2}$ & $\textcolor{red}{-1}$ \\
$\textcolor{blue}{6}$ & $1$ & $0$ & $0$ & $1$ & $1$ & $4$\\
$\textcolor{blue}{7}$ & $1$ & $0$ & $1$ & $2$ & $0$ & $3$\\
$\textcolor{red}{2i}$ & $\textcolor{red}{1}$ & $\textcolor{red}{0}$ & $\textcolor{red}{2}$ & $\textcolor{red}{3}$ & $\textcolor{red}{-1}$ & $\textcolor{red}{2}$ \\
$\textcolor{blue}{8}$ & $1$ & $1$ & $1$ & $1$ & $1$ & $2$\\
$\textcolor{blue}{9}$ & $1$ & $1$ & $2$ & $2$ & $0$ & $1$\\
$\textcolor{blue}{10}$ & $1$ & $2$ & $2$ & $1$ & $1$ & $0$\\

\hline

\end{tabular}
\caption{Number of unlabeled graphs with $N=7$ and $3\leq\mathcal L_1<\mathcal L_2 = \mathcal L_3$. $\mathcal L_1 = 4$ and $\mathcal L_2=\mathcal L_3 = 5$. Invalid entries are highlighted in red.}
\label{tab:ineq_table_L2eL3}
\end{table}

\begin{table}[h!]
\centering
\begin{tabular}{|*{14}{c|}}
\hline
\multicolumn{7}{|c|}{$\mathcal L_1=3$, $\mathcal L_2=3$, $\mathcal L_3=8$} & \multicolumn{7}{c|}{$\mathcal L_1=4$, $\mathcal L_2=4$, $\mathcal L_3=6$} \\
\hline
index & $a$ & $b$ & $c$ & $d$ & $e$ & $f$ & index & $a$ & $b$ & $c$ & $d$ & $e$ & $f$\\
\hline
$\textcolor{red}{1i}$ & $\textcolor{red}{0}$ & $\textcolor{red}{0}$ & $\textcolor{red}{0}$ & $\textcolor{red}{-1}$ & $\textcolor{red}{4}$ & $\textcolor{red}{4}$ & $\textcolor{blue}{1}$ & $0$ & $0$ & $0$ & $1$ & $3$ & $3$\\
$\textcolor{blue}{1}$ & $0$ & $0$ & $1$ & $0$ & $3$ & $3$& $\textcolor{blue}{2}$ & $0$ & $0$ & $1$ & $2$ & $2$ & $2$\\
$\textcolor{blue}{2}$ & $0$ & $0$ & $2$ & $1$ & $2$ & $2$ & $\textcolor{blue}{3}$ & $0$ & $0$ & $2$ & $3$ & $1$ & $1$\\
$\textcolor{blue}{3}$ & $0$ & $0$ & $3$ & $2$ & $1$ & $1$ & $\textcolor{blue}{4}$ & $0$ & $1$ & $0$ & $0$ & $4$ & $2$\\
$\textcolor{red}{2i}$ & $\textcolor{red}{0}$ & $\textcolor{red}{1}$ & $\textcolor{red}{0}$ & $\textcolor{red}{-2}$ & $\textcolor{red}{5}$ & $\textcolor{red}{3}$ & $\textcolor{blue}{5}$ & $0$ & $1$ & $1$ & $1$ & $3$ & $1$\\
$\textcolor{red}{3i}$ & $\textcolor{red}{0}$ & $\textcolor{red}{1}$ & $\textcolor{red}{1}$ & $\textcolor{red}{-1}$ & $\textcolor{red}{4}$ & $\textcolor{red}{2}$ & $\textcolor{blue}{6}$ & $0$ & $1$ & $2$ & $2$ & $2$ & $0$\\
$\textcolor{blue}{4}$ & $0$ & $1$ & $2$ & $0$ & $3$ & $1$ & $\textcolor{red}{1i}$ & $\textcolor{red}{1}$ & $\textcolor{red}{1}$ & $\textcolor{red}{0}$ & $\textcolor{red}{-1}$ & $\textcolor{red}{3}$ & $\textcolor{red}{3}$\\
$\textcolor{blue}{5}$ & $0$ & $1$ & $3$ & $1$ & $2$ & $0$ & $\textcolor{blue}{7}$ & $1$ & $1$ & $1$ & $0$ & $2$ & $2$\\
$\textcolor{red}{4i}$ & $\textcolor{red}{1}$ & $\textcolor{red}{1}$ & $\textcolor{red}{0}$ & $\textcolor{red}{-3}$ & $\textcolor{red}{4}$ & $\textcolor{red}{4}$ & $\textcolor{blue}{8}$ & $1$ & $1$ & $2$ & $1$ & $1$ & $1$\\
$\textcolor{red}{5i}$ & $\textcolor{red}{1}$ & $\textcolor{red}{1}$ & $\textcolor{red}{1}$ & $\textcolor{red}{-2}$ & $\textcolor{red}{3}$ & $\textcolor{red}{3}$ & & & & & & & \\
$\textcolor{red}{6i}$ & $\textcolor{red}{1}$ & $\textcolor{red}{1}$ & $\textcolor{red}{2}$ & $\textcolor{red}{-1}$ & $\textcolor{red}{2}$ & $\textcolor{red}{2}$ & & & & & & & \\
$\textcolor{blue}{6}$ & $1$ & $1$ & $3$ & $0$ & $1$ & $1$ & & & & & & &\\
\hline

\end{tabular}
\caption{Number of unlabeled graphs with $N=7$ and $3\leq\mathcal L_1=\mathcal L_2 < \mathcal L_3$. $\mathcal L_1= \mathcal L_2 = 3$ and $\mathcal L_3 = 8$ to the left, $\mathcal L_1= \mathcal L_2 = 4$ and $\mathcal L_3 = 6$ to the right. Invalid entries are highlighted in red.}
\label{tab:ineq_table_L1eL2}
\end{table}

\section{Three equal degrees}

If $\mathcal L_1 = \mathcal L_2 = \mathcal L_3$, then we have to impose $a\leq b\leq c$ to avoid counting permutation-similar adjacency matrices.
Note that the adjacency matrices with $a=b<c$, $a<b=c$ or $a=b=c$, 
and $\mathcal L_1=\mathcal L_2 = \mathcal L_3$, are of the form 
\renewcommand{\arraystretch}{1.0} 
\begin{equation*}
\begin{pmatrix}
a & d & e \\
d & a & e \\
e & e & c
\end{pmatrix} ,\qquad
\begin{pmatrix}
a & d & d \\
d & b & f \\
d & f & b
\end{pmatrix},\qquad
\begin{pmatrix}
a & d & d \\
d & a & d \\
d & d & a
\end{pmatrix},
\end{equation*}
\renewcommand{\arraystretch}{1.7} 
which are invariant under the permutations of nodes $(1,2)$, $(2,3)$, and $(1,2,3)$, respectively.

\begin{equation}
\begin{split}
|G(3\leq\mathcal L_1=\mathcal L_2=\mathcal L_3)| = \sum_{a=0}^{\left\lfloor \frac{\mathcal L_1-1}{2} \right\rfloor}\sum_{ b=a}^{\left\lfloor \frac{\mathcal L_2-1}{2} \right\rfloor}\sum_{ c=b}^{\left\lfloor \frac{\mathcal L_3-1}{2} \right\rfloor} 1&-\sum_{ a=0}^{\left\lfloor \frac{\mathcal L_1-1}{2} \right\rfloor}\sum_{ b=a}^{\left\lfloor \frac{\mathcal L_2-1}{2} \right\rfloor}\sum_{ c=b}^{\min\left(\left\lfloor \frac{\mathcal L_3-1}{2} \right\rfloor,a+b-\frac{\mathcal L_1+\mathcal L_2-\mathcal L_3}{2}\right)} 1 \\
&-\sum_{ a=0}^{\left\lfloor \frac{\mathcal L_1-1}{2} \right\rfloor}\sum_{ b=a}^{\left\lfloor \frac{\mathcal L_2-1}{2} \right\rfloor}\sum_{ c=\max\left(b,-a+b+\frac{\mathcal L_1-\mathcal L_2+\mathcal L_3}{2}+1\right)}^{\left\lfloor \frac{\mathcal L_3-1}{2} \right\rfloor} 1 \\
&-\sum_{ a=0}^{\left\lfloor \frac{\mathcal L_1-1}{2} \right\rfloor}\sum_{ b=a}^{\left\lfloor \frac{\mathcal L_2-1}{2} \right\rfloor}\sum_{ c=\max\left(b,a-b+\frac{-\mathcal L_1+\mathcal L_2+\mathcal L_3}{2}+1\right)}^{\left\lfloor \frac{\mathcal L_3-1}{2} \right\rfloor} 1\; .
\label{eq:number_ineq_4_3}
\end{split}
\end{equation}

Finally, if we define

\begin{equation}
g_{11}(A):=\frac{A(A+2)(2A-1)}{24}\;,
\label{eq:gs_4}
\end{equation}
then

\begin{equation}
|G(3\leq\mathcal L_1=\mathcal L_2=\mathcal L_3)|=\frac{1}{6}\left\lfloor\frac{\mathcal L_1+1}{2}\right\rfloor \left\lfloor\frac{\mathcal L_1+3}{2}\right\rfloor \left\lfloor\frac{\mathcal L_1+5}{2}\right\rfloor-g_{11}(A_1)\;.
\label{eq:counts_summed_4}
\end{equation}

\subsection{Example with three equal degrees}

\begin{figure}
    \centering
    \includegraphics[width=0.85\linewidth]{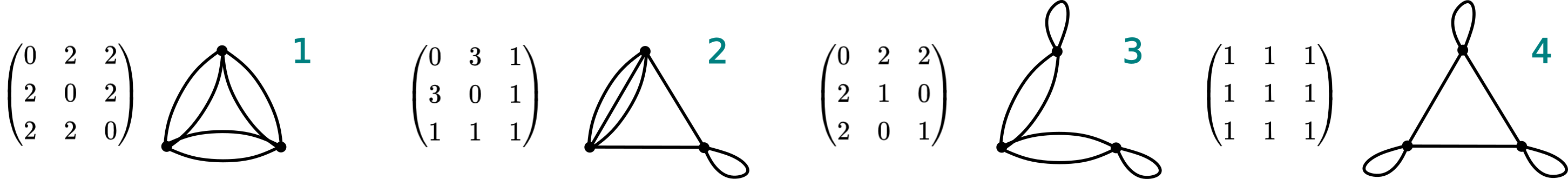}
    \caption{Cconnected unlabeled graphs with $\mathcal L_1=\mathcal L_2 = \mathcal L_3=4$.}
    \label{fig:ineq444}
\end{figure}

\begin{figure}
    \centering
    \includegraphics[width=0.95\linewidth]{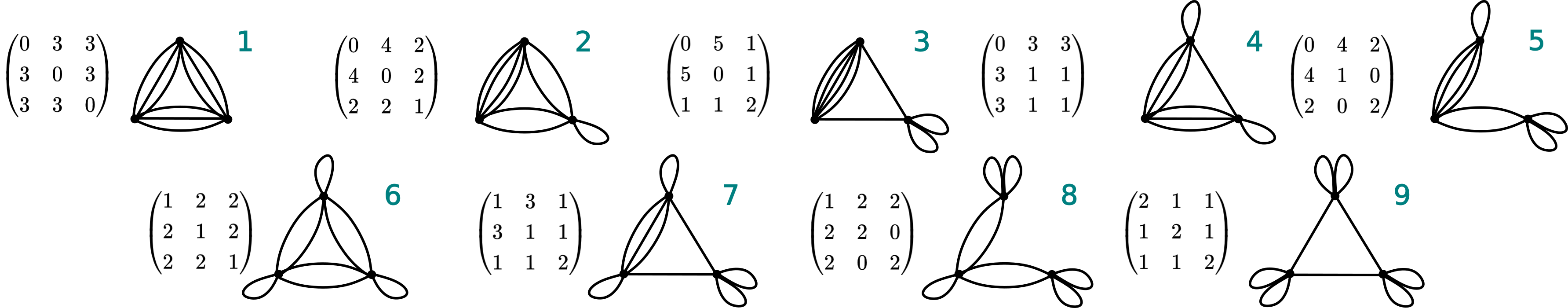}
    \caption{Cconnected unlabeled graphs with $\mathcal L_1=\mathcal L_2 = \mathcal L_3=6$.}
    \label{fig:ineq666}
\end{figure}

Eq. \eqref{eq:num_deg_3eq} states that, for every $N$ at most one combination with $3\leq\mathcal L_1 = \mathcal L_2 = \mathcal L_3$ exists, specifically only if $3|N$. In particular, there are no combinations with $3\leq\mathcal L_1 = \mathcal L_2 = \mathcal L_3$ and $N=7$. For $N=6,9,12$ and $15$, Eq. \eqref{eq:counts_summed_4} gives $|G(4,4,4)|=4$, $|G(6,6,6)|=9$, $|G(8,8,8)|=17$
and $|G(10,10,10)|=28$. Table \ref{tab:ineq_table_L1eL2eL3} and Figs. \ref{fig:ineq444} and \ref{fig:ineq666} report all unlabeled graphs with $\mathcal L_1 = \mathcal L_2 = \mathcal L_3 = 4$ and $\mathcal L_1 = \mathcal L_2 = \mathcal L_3 = 6$.
\begin{table}[h!]
\centering
\begin{tabular}{|*{14}{c|}}
\hline
\multicolumn{7}{|c|}{$\mathcal L_1=4$, $\mathcal L_2=4$, $\mathcal L_3=4$} & \multicolumn{7}{c|}{$\mathcal L_1=6$, $\mathcal L_2=6$, $\mathcal L_3=6$} \\
\hline
index & $a$ & $b$ & $c$ & $d$ & $e$ & $f$ & index & $a$ & $b$ & $c$ & $d$ & $e$ & $f$\\
\hline
$\textcolor{blue}{1}$ & $0$ & $0$ & $0$ & $2$ & $2$ & $2$& $\textcolor{blue}{1}$ & $0$ & $0$ & $0$ & $3$ & $3$ & $3$\\
$\textcolor{blue}{2}$ & $0$ & $0$ & $1$ & $3$ & $1$ & $1$ & $\textcolor{blue}{2}$ & $0$ & $0$ & $1$ & $4$ & $2$ & $2$\\
$\textcolor{blue}{3}$ & $0$ & $1$ & $1$ & $2$ & $2$ & $0$ & $\textcolor{blue}{3}$ & $0$ & $0$ & $2$ & $5$ & $1$ & $1$\\
$\textcolor{blue}{4}$ & $1$ & $1$ & $1$ & $1$ & $1$ & $1$ & $\textcolor{blue}{4}$ & $0$ & $1$ & $1$ & $3$ & $3$ & $1$\\
 & & & & & & & $\textcolor{blue}{5}$ & $0$ & $1$ & $2$ & $4$ & $2$ & $0$\\
 & & & & & & & $\textcolor{red}{1i}$ & $\textcolor{red}{0}$ & $\textcolor{red}{2}$ & $\textcolor{red}{2}$ & $\textcolor{red}{3}$ & $\textcolor{red}{3}$ & $\textcolor{red}{-1}$\\
 & & & & & & & $\textcolor{blue}{6}$ & $1$ & $1$ & $1$ & $2$ & $2$ & $2$\\
 & & & & & & & $\textcolor{blue}{7}$ & $1$ & $1$ & $2$ & $3$ & $1$ & $1$\\
 & & & & & & & $\textcolor{blue}{8}$ & $1$ & $2$ & $2$ & $2$ & $2$ & $0$\\
 & & & & & & & $\textcolor{blue}{9}$ & $2$ & $2$ & $2$ & $1$ & $1$ & $1$\\
\hline

\end{tabular}
\caption{Number of unlabeled graphs with $3\leq\mathcal L_1=\mathcal L_2 = \mathcal L_3$. $\mathcal L_1= \mathcal L_2 = \mathcal L_3 = 4$ ($N=6$) to the left and $\mathcal L_1= \mathcal L_2 = \mathcal L_3 = 6$ ($N=9$) to the right. Invalid entries are highlighted in red.}
\label{tab:ineq_table_L1eL2eL3}
\end{table}

\section{Conclusions}

In this work, we counted all unlabeled, undirected, connected (and traversable) multigraphs with 3 nodes of degree $\mathcal L_1\geq 3$, $\mathcal L_2\geq 3$ and $\mathcal L_3\geq 3$, for any value of $\mathcal L_1$, $\mathcal L_2$ and $\mathcal L_3$. This study complements and extends our previous results~\cite{Bonato2024,Bonato2024pre,Bonato2025preprint}, and, combined with the statistical mechanics of polymer networks, yields analytical tools to search for preferred patterns in folded chromatin networks.
Inspired by polymer physics and chromatin organisation, we focused on traversable graphs with node degrees greater than 3. This is the simplest case in which the graphs are not linear, which is presumably relevant in chromatin networks under confinement - i.e. not in the dilute regime.
Nevertheless, the described procedure could potentially be extended to include different structural constraints, such as disconnected graphs, therefore providing enumeration techniques which could be useful in other contexts.
Although analytically enumerating unlabeled graphs with many nodes remains very challenging due to the increasing complexity (symmetries) of the problem, it would be interesting to explore how far the content of this work could be generalised to include, at least partially, graphs with more than three nodes.



\section*{Acknowledgments}
I thank Prof. Davide Marenduzzo, Prof. Enzo Orlandini, and Prof. Sergey Kitaev for their precious suggestions.

\bibliographystyle{apsrev4-1}
\bibliography{references_inequivalent_3V}

\appendix

\section{Summation to polynomial expressions}
\label{appendix:A}

\subsection{ $\mathcal L_1 \neq \mathcal L_2 \neq \mathcal L_3$}

In Section~\eqref{eq:conn_eqs_3}, we informed that the number of unlabeled connected (and traversable) graphs with degrees $\mathcal L_1$, $\mathcal L_2$, $\mathcal L_3$, and $\mathcal L_1 \neq \mathcal L_2 \neq \mathcal L_3$ is given by Eq.~\eqref{eq:number_ineq_1_3}. Here, we report how Eq.~\eqref{eq:number_ineq_1_3} can be transformed into a polynomial expression.

By summing up the first term in Eq.~\eqref{eq:number_ineq_1_3}, summing over $c$ the others and relabeling, we have 

\begin{equation}
\begin{split}
|G(3\leq\mathcal L_1<\mathcal L_2<\mathcal L_3)|= \prod_{i=1}^3 \left\lfloor \frac{\mathcal L_i+1}{2} \right\rfloor - &\sum_{a=0}^{\left\lfloor \frac{\mathcal L_1-1}{2} \right\rfloor}\sum_{b=0}^{\left\lfloor \frac{\mathcal L_2-1}{2} \right\rfloor} \max\left(0,\min\left( \left\lfloor \frac{\mathcal L_3-1}{2} \right\rfloor,a+b-\frac{\mathcal L_1+\mathcal L_2-\mathcal L_3}{2}-1\right) +1\right) \\
- &\sum_{a=0}^{\left\lfloor \frac{\mathcal L_1-1}{2} \right\rfloor}\sum_{c=0}^{\left\lfloor \frac{\mathcal L_3-1}{2} \right\rfloor}\max\left(0,\min\left(\left\lfloor \frac{\mathcal L_2-1}{2} \right\rfloor,a+c-\frac{\mathcal L_1-\mathcal L_2+\mathcal L_3}{2}-1\right)+1\right) \\ 
- &\sum_{b=0}^{\left\lfloor \frac{\mathcal L_2-1}{2} \right\rfloor}\sum_{c=0}^{\left\lfloor \frac{\mathcal L_3-1}{2} \right\rfloor} \max\left(0, \min\left( \left\lfloor \frac{\mathcal L_1-1}{2} \right\rfloor,  b+c-\frac{-\mathcal L_1+\mathcal L_2+\mathcal L_3}{2}-1\right)+1\right) \\
= \prod_{i=1}^3 \left\lfloor \frac{\mathcal L_i+1}{2} \right\rfloor - &\sum_{i=1}^{3}\sum_{a=0}^{A_i}\sum_{b=0}^{B_i} \max\left(0,\min\left( C_i,a+b-K_i-1\right) +1\right)\;,
\end{split}
\label{eq:number_ineq_app_1}
\end{equation}
where the coefficients $A_i$, $B_i$, $C_i$, and $K_i$ are reported in Table~\ref{tab:coeffs}.
\renewcommand{\arraystretch}{1.7} 
\begin{table}[h!]
\renewcommand{\thetable}{\ref{tab:coeffs}}
\centering
\begin{tabular}{|c|c|c|c|c|}
\hline
$i$ & $A_i$ & $B_i$ & $C_i$ & $K_i$ \\
\hline
$1$ & $\left\lfloor \frac{\mathcal L_1-1}{2} \right\rfloor$ & $\left\lfloor \frac{\mathcal L_2-1}{2} \right\rfloor$ & $\left\lfloor \frac{\mathcal L_3-1}{2} \right\rfloor$ & $\frac{\mathcal L_1+\mathcal L_2-\mathcal L_3}{2}$ \\
$2$ & $\left\lfloor \frac{\mathcal L_1-1}{2} \right\rfloor$ & $\left\lfloor \frac{\mathcal L_3-1}{2} \right\rfloor$ & $\left\lfloor \frac{\mathcal L_2-1}{2} \right\rfloor$ & $\frac{\mathcal L_1-\mathcal L_2+\mathcal L_3}{2}$ \\
$3$ & $\left\lfloor \frac{\mathcal L_2-1}{2} \right\rfloor$ & $\left\lfloor \frac{\mathcal L_3-1}{2} \right\rfloor$ & $\left\lfloor \frac{\mathcal L_1-1}{2} \right\rfloor$ & $\frac{-\mathcal L_1+\mathcal L_2+\mathcal L_3}{2}$ \\
\hline
\end{tabular}
\caption{Coefficients in Eq.~\eqref{eq:number_ineq_app_1}.}
\end{table}

The sum

\begin{equation*}
F=\sum_{a=0}^{A}\sum_{b=0}^{B} \max\left(0,\min\left( C,a+b-K-1\right) +1\right)
\end{equation*}
\noindent can be expressed as a polynomial by adopting the change of variable $a,b\to a,s=a+b$ 
\begin{equation}
\begin{split}
F&=\sum_{s=K+1}^{A+B}\sum_{a=\max(0,s-B)}^{\min(s,A)} \min\left( C,s-K-1\right) +1 \\
&= \sum_{s=K+1}^{A+B} \left(\min\left( C,s-K-1\right) +1\right)\left(\max\left(0,\min(s,A)-\max(0,s-B)+1\right)\right)
\end{split}
\end{equation}
and by splitting the sum over $s$ as follows

\begin{equation}
\begin{split}
F=&\sum_{s=\max(0,K+1)}^{\min(\min(A,B),K+1+C)} (s+1)(s-K) + \sum_{s=\max(K+1,\min(A,B)+1)}^{\min(\max(A,B),K+1+C)} (\min(A,B)+1)(s-K) \\
+&\sum_{s=\max(K+1,\max(A,B)+1)}^{\min(A+B,K+1+C)} (A+B-s+1)(s-K) + \sum_{s=K+2+C}^{\min(\min(A,B),A+B)} (s+1)(C+1) \\
+ &\sum_{s=\max(K+2+C,\min(A,B)+1)}^{\min(\max(A,B),A+B)} (\min(A,B)+1)(C+1) +\sum_{s=\max(K+2+C,\max(A,B)+1)}^{A+B} (A+B-s+1)(C+1)\; . \\
\end{split}
\label{eq:F}
\end{equation}

To sum up Eq.~\eqref{eq:F} (where, with respect to Eq.~\eqref{eq:number_ineq_2_3}, we dropped the index $i$ for convenience), we recall that $\mathcal L_i$, and consequently $A_i$ $B_i$ and $C_i$ are ordered, specifically $3\leq \mathcal L_1 < \mathcal L_2 < \mathcal L_3$, $1\leq A_1 \leq B_1\leq C_1$, $1\leq A_2 \leq C_2\leq B_2$ and $1\leq C_3 \leq A_3\leq B_3$.
In particular, the ranges of the 6 sums in Eq.~\eqref{eq:F}, after reintroducing the index $i$, are reported in Table~\ref{tab:ranges}.
\begin{table}[h!]
\centering
\begin{tabular}{|c|c|c|c|c|c|c|}
\hline
$i$ & sum 1 & sum 2 & sum 3 & sum 4 & sum 5 & sum 6\\
\hline
$1$ & 
$\begin{aligned} [K_1+1,A_1];&\; \mathcal L_1 + \mathcal L_2 + 1\geq\mathcal L_3 \geq \mathcal L_2+3 \\
[0,A_1];&\; \mathcal L_1 + \mathcal L_2 + 1\geq\mathcal L_3\end{aligned}$
& $[A_1+1,B_1]$;\; $A_1<B_1$ & $[B_1+1,A_1+B_1]$ & empty & empty & empty\\
$2$ & empty & $[K_2+1,B_2]$;\; $\mathcal L_2 \geq \mathcal L_1+3$ & $[B_2+1,A_2+B_2]$ & empty & empty & empty\\
$3$ & empty & empty & $[K_3+1,A_3+B_3]$ & empty & empty & empty\\
\hline
\end{tabular}
\caption{Ranges of the sums in Eq.~\eqref{eq:F} applied to Eq.~\eqref{eq:number_ineq_app_1}.}
\label{tab:ranges}
\end{table}

Finally, if we define $g_1(A,B,K)$, $g_2(A,B,K)$, $g_3(A,B,K)$, $g_4(A,B,K)$, $g_5(A,B,K)$ and $g_6(A,B,K)$ as in Eq.~\eqref{eq:gs}, we obtain Eq.~\eqref{eq:counts_summed}.

\subsection{$\mathcal L_1 <\mathcal L_2 = \mathcal L_3$ and $\mathcal L_1 =\mathcal L_2 < \mathcal L_3$}

In this and in the next subsection we report additional details on the summation of Eqs. \eqref{eq:number_ineq_2_3}, \eqref{eq:number_ineq_3_3} and \eqref{eq:number_ineq_4_3} for the cases $\mathcal L_1 <\mathcal L_2 = \mathcal L_3$, $\mathcal L_1 =\mathcal L_2 < \mathcal L_3$,
 and $\mathcal L_1 =\mathcal L_2 = \mathcal L_3$.
\begin{equation}
\begin{split}
&\sum_{ a=0}^{\left\lfloor \frac{\mathcal L_1-1}{2} \right\rfloor}\sum_{ b=0}^{\left\lfloor \frac{\mathcal L_2-1}{2} \right\rfloor}\sum_{ c=b}^{\min\left(\left\lfloor \frac{\mathcal L_3-1}{2} \right\rfloor,a+b-\frac{\mathcal L_1+\mathcal L_2-\mathcal L_3}{2}\right)} 1 = 0\\
&\sum_{ a=0}^{\left\lfloor \frac{\mathcal L_1-1}{2} \right\rfloor}\sum_{ b=0}^{\left\lfloor \frac{\mathcal L_2-1}{2} \right\rfloor}\sum_{ c=\max\left(b,-a+b+\frac{\mathcal L_1-\mathcal L_2+\mathcal L_3}{2}+1\right)}^{\left\lfloor \frac{\mathcal L_3-1}{2} \right\rfloor} 1 = g_7(A_1,B_1)\\
&\sum_{ a=0}^{\left\lfloor \frac{\mathcal L_1-1}{2} \right\rfloor}\sum_{ b=0}^{\left\lfloor \frac{\mathcal L_2-1}{2} \right\rfloor}\sum_{ c=\max\left(b,a-b+\frac{-\mathcal L_1+\mathcal L_2+\mathcal L_3}{2}+1\right)}^{\left\lfloor \frac{\mathcal L_3-1}{2} \right\rfloor} 1\; = g_8\left(A_1,B_1,
p=\begin{cases}
2 &\text{if}\;2 |\mathcal L_2\\
1 &\text{otherwise}    
\end{cases}\right),
\end{split}
\end{equation}

\begin{equation}
\begin{split}
&\sum_{ a=0}^{\left\lfloor \frac{\mathcal L_1-1}{2} \right\rfloor}\sum_{ b=a}^{\left\lfloor \frac{\mathcal L_2-1}{2} \right\rfloor}\sum_{ c=0}^{\min\left(\left\lfloor \frac{\mathcal L_3-1}{2} \right\rfloor,a+b-\frac{\mathcal L_1+\mathcal L_2-\mathcal L_3}{2}\right)} 1 = g_{9}\left(A_1,C_1,p=\begin{cases}
2 &\text{if}\;2 |\mathcal L_1\\
1 & \text{otherwise}    
\end{cases}\right)\\
&\sum_{ a=0}^{\left\lfloor \frac{\mathcal L_1-1}{2} \right\rfloor}\sum_{ b=a}^{\left\lfloor \frac{\mathcal L_2-1}{2} \right\rfloor}\sum_{ c=\max\left(0,-a+b+\frac{\mathcal L_1-\mathcal L_2+\mathcal L_3}{2}+1\right)}^{\left\lfloor \frac{\mathcal L_3-1}{2} \right\rfloor} 1 =0 \\
&\sum_{ a=0}^{\left\lfloor \frac{\mathcal L_1-1}{2} \right\rfloor}\sum_{ b=a}^{\left\lfloor \frac{\mathcal L_2-1}{2} \right\rfloor}\sum_{ c=\max\left(0,a-b+\frac{-\mathcal L_1+\mathcal L_2+\mathcal L_3}{2}+1\right)}^{\left\lfloor \frac{\mathcal L_3-1}{2} \right\rfloor} 1 = g_{10}(A_1),
\end{split}
\end{equation}
\noindent where $g_7$, $g_8$, $g_9$ and $g_{10}$ are defined in Eqs. \eqref{eq:gs_2} and \eqref{eq:gs_3}.

\subsection{$\mathcal L_1 =\mathcal L_2 = \mathcal L_3$}

\begin{equation}
\begin{split}
&\sum_{ a=0}^{\left\lfloor \frac{\mathcal L_1-1}{2} \right\rfloor}\sum_{ b=a}^{\left\lfloor \frac{\mathcal L_2-1}{2} \right\rfloor}\sum_{ c=b}^{\min\left(\left\lfloor \frac{\mathcal L_3-1}{2} \right\rfloor,a+b-\frac{\mathcal L_1+\mathcal L_2-\mathcal L_3}{2}\right)} 1 = 0\\
&\sum_{ a=0}^{\left\lfloor \frac{\mathcal L_1-1}{2} \right\rfloor}\sum_{ b=a}^{\left\lfloor \frac{\mathcal L_2-1}{2} \right\rfloor}\sum_{ c=\max\left(b,-a+b+\frac{\mathcal L_1-\mathcal L_2+\mathcal L_3}{2}+1\right)}^{\left\lfloor \frac{\mathcal L_3-1}{2} \right\rfloor} 1 =0 \\
&\sum_{ a=0}^{\left\lfloor \frac{\mathcal L_1-1}{2} \right\rfloor}\sum_{ b=a}^{\left\lfloor \frac{\mathcal L_2-1}{2} \right\rfloor}\sum_{ c=\max\left(b,a-b+\frac{-\mathcal L_1+\mathcal L_2+\mathcal L_3}{2}+1\right)}^{\left\lfloor \frac{\mathcal L_3-1}{2} \right\rfloor} 1 = g_{11}(A_1) ,
\end{split}
\end{equation}
\noindent where $g_{11}$ is defined in Eq. \eqref{eq:gs_4}.

\begin{table}[t!]
\centering
\begin{tabular}{|*{10}{c|}}
\hline
\multicolumn{2}{|c|}{$N=6$} & \multicolumn{2}{|c|}{$N=7$} & \multicolumn{2}{|c|}{$N=8$}& \multicolumn{2}{|c|}{$N=9$}& \multicolumn{2}{|c|}{$N=10$}\\
\hline
$\mathcal L_1$ $\mathcal L_2$ $\mathcal L_3$ & $|G_3|$ & $\mathcal L_1$ $\mathcal L_2$ $\mathcal L_3$ & $|G_3|$ & $\mathcal L_1$ $\mathcal L_2$ $\mathcal L_3$ & $|G_3|$ & $\mathcal L_1$ $\mathcal L_2$ $\mathcal L_3$ & $|G_3|$ & $\mathcal L_1$ $\mathcal L_2$ $\mathcal L_3$ & $|G_3|$\\
\hline
$3$ $4$ $5$ & $10$ & $3$ $4$ $7$ & $11$ & $3$ $4$ $9$ & $11$ & $3$ $4$ $11$ & $11$ & $3$ $4$ $13$ & $11$\\
 & & $3$ $5$ $6$ & $13$ & $3$ $5$ $8$ & $14$ & $3$ $5$ $10$ & $14$ & $3$ $5$ $12$ & $14$\\
  & & & & $3$ $6$ $7$ & $16$ & $3$ $6$ $9$ & $17$ & $3$ $6$ $11$ & $17$\\
  &  & & & $4$ $5$ $7$ & $18$ & $3$ $7$ $8$ & $19$ & $3$ $7$ $10$ & $20$\\
& & & & & & $4$ $5$ $9$ & $19$ & $3$ $8$ $9$ & $22$\\
& & & & & & $4$ $6$ $8$ & $19$ & $4$ $5$ $11$ & $19$\\
& & & & & & $5$ $6$ $7$ & $27$ & $4$ $6$ $10$ & $20$\\
& & & & & & & & $4$ $7$ $9$ & $26$\\
& & & & & & & & $5$ $6$ $9$ & $30$\\
& & & & & & & & $5$ $7$ $8$ & $33$\\

\hline

\end{tabular}
\caption{Number of unlabeled graphs with $N=6,7,8,9$ and $10$ edges, and $3\leq\mathcal L_1<\mathcal L_2<\mathcal L_3$.}
\label{tab:ineq_table_6to10}
\end{table}

\section{Number of unlabeled graphs for different number of edges and degrees}
\label{appendix:B}

In this section, we report the number of unlabeled graphs for different number of edges $N = 6,7,8,9,10,15$ and gven values of $\mathcal L_1$, $\mathcal L_2$ and $\mathcal L_3$. We distinguish between $\mathcal L_1 <\mathcal L_2 <\mathcal L_3$ (tables \ref{tab:ineq_table_6to10} and \ref{tab:ineq_table_15})
and $\mathcal L_1 <\mathcal L_2 = \mathcal L_3$ or $\mathcal L_1 =\mathcal L_2 <\mathcal L_3$ (Table \ref{tab:ineq_table_6to15}).

\begin{table}[h!]
\centering
\begin{tabular}{|*{10}{c|}}
\hline
\multicolumn{10}{|c|}{$N=15$} \\
\hline
$\mathcal L_1$ $\mathcal L_2$ $\mathcal L_3$ & $|G_3|$ & $\mathcal L_1$ $\mathcal L_2$ $\mathcal L_3$ & $|G_3|$ & $\mathcal L_1$ $\mathcal L_2$ $\mathcal L_3$ & $|G_3|$ & $\mathcal L_1$ $\mathcal L_2$ $\mathcal L_3$ & $|G_3|$ & $\mathcal L_1$ $\mathcal L_2$ $\mathcal L_3$ & $|G_3|$\\
\hline
$3$ $4$ $23$ & $11$ & $3$ $12$ $15$ & $35$ & $4$ $11$ $15$ & $43$ & $5$ $11$ $14$ & $60$ & $7$ $8$ $15$ & $66$\\
$3$ $5$ $22$ & $14$ & $3$ $13$ $14$ & $37$ & $4$ $12$ $14$ & $43$ & $5$ $12$ $13$ & $63$ & $7$ $9$ $14$ & $75$\\
$3$ $6$ $21$ & $17$ & $4$ $5$ $21$ & $19$ & $5$ $6$ $19$ & $31$ & $6$ $7$ $17$ & $46$ & $7$ $10$ $13$ & $82$\\
$3$ $7$ $20$ & $20$ & $4$ $6$ $20$ & $20$ & $5$ $7$ $18$ & $37$ & $6$ $8$ $16$ & $49$ & $7$ $11$ $12$ & $86$\\
$3$ $8$ $19$ & $23$ & $4$ $7$ $19$ & $27$ & $5$ $8$ $17$ & $43$ & $6$ $9$ $15$ & $61$ & $8$ $9$ $13$ & $86$\\
$3$ $9$ $18$ & $26$ & $4$ $8$ $18$ & $28$ & $5$ $9$ $16$ & $49$ & $6$ $10$ $14$ & $63$ & $8$ $10$ $12$ & $86$\\
$3$ $10$ $17$& $29$ & $4$ $9$ $17$ & $35$ & $5$ $10$ $15$ & $55$ & $6$ $11$ $13$ & $72$ & $9$ $10$ $11$ & $100$\\
$3$ $11$ $16$& $32$ & $4$ $10$ $16$ & $36$ & & & & & & \\

\hline
\end{tabular}
\caption{Number of unlabeled graphs with $N=15$ edges,and $3\leq\mathcal L_1<\mathcal L_2<\mathcal L_3$.}
\label{tab:ineq_table_15}
\end{table}

\begin{table}[h!]
\centering
\resizebox{\textwidth}{!}{
\centering
\begin{tabular}{|*{12}{c|}}
\hline
\multicolumn{2}{|c|}{$N=6$} & \multicolumn{2}{|c|}{$N=7$} & \multicolumn{2}{|c|}{$N=8$}& \multicolumn{2}{|c|}{$N=9$}& \multicolumn{2}{|c|}{$N=10$}& \multicolumn{2}{|c|}{$N=15$}\\
\hline
$\mathcal L_1$ $\mathcal L_2$ $\mathcal L_3$ & $|G_3|$ & $\mathcal L_1$ $\mathcal L_2$ $\mathcal L_3$ & $|G_3|$ & $\mathcal L_1$ $\mathcal L_2$ $\mathcal L_3$ & $|G_3|$ & $\mathcal L_1$ $\mathcal L_2$ $\mathcal L_3$ & $|G_3|$ & $\mathcal L_1$ $\mathcal L_2$ $\mathcal L_3$ & $|G_3|$ &$\mathcal L_1$ $\mathcal L_2$ $\mathcal L_3$ & $|G_3|$\\
\hline
$3$ $3$ $6$ & $6$ & $3$ $3$ $8$ & $6$ & $3$ $3$ $10$ & $6$ & $3$ $3$ $12$ & $6$ & $3$ $3$ $14$ & $6$ & $3$ $3$ $24$ & $6$\\
 & & $4$ $4$ $6$& $8$& $4$ $4$ $8$ & $9$ & $4$ $4$ $10$ & $9$ & $4$ $4$ $12$ & $9$ & $4$ $4$ $22$ & $9$\\
& & $4$ $5$ $5$ & $10$ & $4$ $6$ $6$ & $11$ & $4$ $7$ $7$ & $15$ & $4$ $8$ $8$ & $16$ & $4$ $13$ $13$ & $30$\\
  & & &  & $5$ $5$ $6$ & $14$ & $5$ $5$ $8$ & $16$ & $5$ $5$ $10$ & $17$ & $5$ $5$ $20$ & $17$\\
& & & & & & & & $6$ $6$ $8$ & $20$ & $6$ $6$ $18$ & $23$\\
& & & & & & & & $6$ $7$ $7$ & $23$ & $6$ $12$ $12$ & $43$\\
& & & & & & & & & & $7$ $7$ $16$ & $36$\\
& & & & & & & & & & $8$ $8$ $14$ & $45$\\
& & & & & & & & & & $8$ $11$ $11$ & $57$\\
& & & & & & & & & & $9$ $9$ $12$ & $58$\\

\hline

\end{tabular}
}
\caption{Number of unlabeled graphs with $N=6,7,8,9$,$10$ and $15$ edges with $3\leq\mathcal L_1=\mathcal L_2<\mathcal L_3$, and $3\leq\mathcal L_1<\mathcal L_2=\mathcal L_3$.}
\label{tab:ineq_table_6to15}
\end{table}

\end{document}